\documentclass[a4paper]{article}
\usepackage[left=3cm,right=3cm,top=2cm,bottom=2.5cm]{geometry}
\usepackage{times}
\usepackage{authblk}
\usepackage{graphicx}
\usepackage{amsmath, amssymb}
\usepackage{enumitem} 
\usepackage{subcaption} 

\usepackage{booktabs}
\usepackage{color}
\usepackage[usenames,dvipsnames]{xcolor}
\usepackage[round]{natbib}

\usepackage{algorithm}%
\usepackage{algpseudocode}%

\usepackage[]{graphicx}
\chardef\bslash=`\\ 

\numberwithin{equation}{section} 
\usepackage{hyperref} 

\begin{document}

\title{Estimating Zero-inflated Negative Binomial GAMLSS via \\
a Balanced Gradient Boosting Approach \\
with an Application to Antenatal Care Data from Nigeria} 
\date{}
\author[$1$,*]{Alexandra Daub}
\author[$1$]{Elisabeth Bergherr}
\affil[$1$]{ \small Chair of Spatial Data Science and Statistical Learning, Georg-August-Universit\"at G\"ottingen, \hspace*{1em} \newline
\hspace*{-17.6em} Platz der G\"ottinger Sieben 3, 37073 G\"ottingen, Germany} 
\affil[ ]{\hspace*{-20em} $^*$ \small Correspondence: alexandra.daub@uni-goettingen.de \hspace*{20em}}
\maketitle  
\vspace*{-3em}

\begin{abstract}
\footnotesize{Statistical boosting algorithms are renowned for their intrinsic variable selection and enhanced predictive performance compared to classical statistical methods, making them especially useful for complex models such as generalized additive models for location scale and shape (GAMLSS).
Boosting this model class can suffer from imbalanced updates across the distribution parameters as well as long computation times.
Shrunk optimal step lengths have been shown to address these issues. 
To examine the influence of socio-economic factors on the distribution of the number of antenatal care visits in Nigeria, we generalize boosting of GAMLSS with shrunk optimal step lengths to base-learners beyond simple linear models and to a more complex response variable distribution.
In an extensive simulation study and in the application we demonstrate that shrunk optimal step lengths yield a more balanced regularization of the overall model and enhance computational efficiency across diverse settings, in particular in the presence of base-learners penalizing the size of the fit. \\[0.5em] 


}
\end{abstract}
\section{Introduction}

Generalized additive models for location, scale and shape \citep[GAMLSS;][]{Rigby2005} have received increasing attention in recent years \citep{Kneib2013, Klein2024}, being able to model parameters beyond the mean and to incorporate a variety of effects. 
They extend generalized additive models \citep[GAMs;][]{Hastie1990} in that all distribution parameters are modeled based on an additive predictor instead of one.
Due to their ability to model the response variable distribution in a very flexible manner, GAMLSS are used in a variety of  fields. 
In particular, GAMLSS are e.g. recommended for estimating child growth curves, where they are used to model the distribution of characteristics like height and weight as a function of age \citep{WHOchildgrowth2006}.
Beyond biostatistical settings, GAMLSS have also been applied in fields such as finance and for modeling weather and climate-related measurements \citep{Ganegoda2012, Villarini2010}.

While typically estimated using penalized maximum likelihood, alternative estimation techniques have emerged to take advantage of their particular strengths, one of which is statistical boosting \citep{Mayr2012, Thomas2018}. 
In statistical boosting algorithms \citep{Friedman2001, Buehlmann2003}, regression models with a low degree of freedom are iteratively fitted to the current (pseudo-) residuals,
resulting in an ensemble of weak learners as final model. 
Stopping the iterative updating scheme early allows the method to perform variable selection as well as coefficient shrinkage, 
often enhancing predictive performance compared to traditional methods.
Because of this ability to yield sparse models, 
statistical boosting algorithms are particularly advantageous for complex model classes such as GAMLSS.

The increased model complexity, however, also poses challenges when boosting GAMLSS.
While in general the non-cyclical algorithm introduced by \cite{Thomas2018} is favorable, it has been shown to be prone to imbalanced predictor updates and long computation times \citep{Zhang2022, Daub2025}.
These issues arise when the potential updates of the different submodels are on different scales causing the selection procedure to favor updates of certain submodels over others.
They originate from structural differences in the sizes of negative gradients (hence reflecting the structure of the respective likelihood) and, for example, occur in Gaussian location and scale model with a large variance \citep{Zhang2022}.
One way to address such imbalances between submodels is to apply adaptive rather than fixed step lengths \citep{Zhang2022, Daub2025}, whereby the adaptive step lengths compensate for otherwise small update sizes and thus enable a fairer update selection. 
A more detailed explanation of this topic is provided in section~\ref{daub:section_methods}. 

To date, adaptive step lengths have only been applied for 
boosting GAMLSS with two-parameter response variable distributions and simple linear base-learners.
\cite{Zhang2022} consider a Gaussian location and scale model with simple linear base-learners and propose using shrunk optimal step lengths to address the balancing issue.
\cite{Daub2025} investigate a different type of adaptive step length, in addition to shrunk optimal step lengths, for negative binomial and Weibull distributed response variables.
In this work, we generalize this approach with respect to two aspects. 
First and most importantly, a wider variety of base-learner types is considered, specifically for non-linear, categorical and spatial effects, extending beyond simple linear base-learners. 
Second, we apply the boosting algorithm to a zero-inflated negative binomial model for location, scale and shape (ZINB-GAMLSS), 
which has a more complex response variable distribution with substantial dependencies among its parameters. 

The motivation of this extension is to enable a more sophisticated investigation of the relationship between antenatal care utilization in Nigeria and socio-economic, demographic and contextual characteristics of the mother.
With 560 deaths per 100,000 live births reported in 2013, compared to 16 deaths per 100,000 live births in OECD member countries \citep{WHOmaternalmortality2014}, Nigeria is among the countries with the highest maternal mortality worldwide.
Moreover, the country has an insufficient antenatal care utilization: 
69.8\% of the women receive fewer than the recommended minimum number of seven antenatal care visits for pregnancies without complications and 34.2\% report having received no antenatal care at all \citep{DHSNigeria2013, WHOpregnancyrec2016}. 
Since antenatal care enables the identification of pregnancy-related and delivery-related risks and offers timely and appropriate interventions, strengthening the antenatal care utilization is regarded a key strategy for reducing maternal mortality \citep{WHOpregnancyrec2016}. 
To investigate the effects on antenatal care utilization, we consider data from the 2013 Nigeria Demographic and Health Survey \citep{DHSNigeria2013, Gayawan2016}. 
In addition to the number of antenatal care visits, it comprises inter alia data on the mother's education, her age at birth, access to mass media and the region of residence.

From a statistical perspective, the number of antenatal care visits is characterized by excess zeros and a pronounced overdispersion. 
As moreover receiving no antenatal care is especially problematic and the factors associated with this outcome are therefore of great interest, a zero-inflated negative binomial model for location, scale and shape is considered. 
This model framework allows to distinguish between the covariate effects on the conditional mean and those on the zero-inflation probability,
and accommodates a variety of effect types,
which is advantageous given this presence of continuous, categorical and spatial covariates.
In addition to the main effects, interaction terms are included to assess whether specific covariate combinations are particularly problematic.  
As this gives rise to 83 potential effects for each distribution parameter, an estimation method with intrinsic variable selection and regularization is required, where 
statistical boosting is a prominent choice. 
When applying non-cyclical boosting with fixed step lengths, we encountered the mentioned problems of long run times and imbalanced predictor updates.

The remainder of this article is structured as follows. Section~\ref{daub:section_methods} introduces GAMLSS as well as component-wise gradient boosting and expands on how using shrunk optimal step lengths improves the balancedness of the estimated model.
In addition, the special properties of shrunk optimal step lengths when combined with penalized base-learners as well as an alternative numerical approach to obtain optimal step lengths for a ZINB-GAMLSS are presented.
Section~\ref{daub:section_simulations} contains simulation results for a Gaussian location and scale model with different effect types as well as the zero-inflated negative binomial model. 
In section~\ref{daub:section_applications} the results from modeling the number of antenatal care visits in Nigeria are presented and section~\ref{daub:section_conclusion} concludes with an overview of our findings and a discussion. 
\section{Boosting Generalized Additive Models for Location, Scale and Shape}
\label{daub:section_methods}

\subsection{Generalized Additive Models for Location, Scale and Shape}

In GAMLSS, the response variable $Y$ is assumed to follow a distribution $\mathcal{D}$ whose parameters $\theta_1, ..., \theta_K$ all depend on covariates $\boldsymbol{X} = \left(X_1, ..., X_J \right)^\top$ or suitable subsets \citep{Rigby2005}. 
To account for the parameter space restrictions of $\mathcal{D}$, each distribution parameter $\theta_k, k \in \{1,...,K\}$, is related to a structured additive predictor $\eta_{\theta_k}$ through a bijective link function $g_{\theta_k}$, such that $g_{\theta_k}(\theta_k)=\eta_{\theta_k}$ \citep{Fahrmeir2013}.
For given covariate information $\boldsymbol{X}=\boldsymbol{x}$, the conditional distribution of Y is specified as
\begin{gather*}
Y \mid \boldsymbol{X} = \boldsymbol{x} \sim \mathcal{D}(\theta_1(\boldsymbol{x}), ..., \theta_K(\boldsymbol{x})) \quad \text{with} \\
g_{\theta_k}(\theta_k(\boldsymbol{x})) = \eta_{\theta_k}(\boldsymbol{x}) = \beta_{0,\theta_k} + \sum_{j=1}^{J} f_{j,\theta_k}(x_j) , \, k \in \{1, ..., K\} ,
\end{gather*}
where $f_{j,\theta_k}$ is the assumed type of effect of covariate $X_j, j \in \{1, ..., J\}$, on predictor $\eta_{\theta_k}$. 
Motivated by the application, the effect types we will focus on in this work are:
\begin{itemize}
\setlength\itemsep{-3pt}
\item linear effects, i.e. $f(x_j) = x_j \beta_{j,\theta_k}$ 
\item non-linear effects modeled via penalized splines with B-spline basis of degree three and a second order difference penalty \citep{Eilers1996}
\item categorical effects modeled as joint dummy-coded linear effect 
\item discrete spatial effects represented by a Markov random field with a first-order neighborhood structure \citep{Rue2005} 
\end{itemize}

For a zero-inflated negative binomial response variable for example, the three distribution parameters that are modeled are $\theta_1 = \mu$, $\theta_2 = \alpha$ and $\theta_3 = \pi$, where $\pi$ represents the probability that the response variable $Y$ is generated by a zero process, while $\mu$ and $\alpha$ correspond to the location and scale parameters of the negative binomial distribution that generates $Y$ otherwise.
Using the canonical link functions $g_\mu = \ln$, $g_\alpha = \ln$ and $g_\pi = \text{logit}$, 
the zero-inflated negative binomial model for location, scale and shape (ZINB-GAMLSS) is 
\begin{gather*}
Y \mid \boldsymbol{X}=\boldsymbol{x} \sim \textit{ZINB}(\mu(\boldsymbol{x}), \alpha(\boldsymbol{x}), \pi(\boldsymbol{x})) \quad \text{with} \\
\begin{aligned}
\mu(\boldsymbol{x}) &= \exp\left(\beta_{0,\mu} + \sum_{j=1}^{J} f_{j,\mu}(x_j)\right) \\
\alpha(\boldsymbol{x}) &= \exp\left(\beta_{0,\alpha} + \sum_{j=1}^{J} f_{j,\alpha}(x_j)\right) \\
\pi(\boldsymbol{x})& = \text{logit}^{-1}\left(\beta_{0,\pi} + \sum_{j=1}^{J} f_{j,\pi}(x_j)\right).
\end{aligned}
\end{gather*}

\subsection{Model-based Boosting}

As part of the statistical learning toolbox, model-based boosting arose from the field of machine learning \citep{Freund1996} but gained much attention in the statistical community \citep{Hastie2009} as it can be applied to estimate statistical models \citep{Friedman2000, Friedman2001}. 
Advantages of boosting compared to classical statistical methods include its applicability to high-dimensional data and an intrinsic variable selection \citep{Buehlmann2006, Mayr2014b}. 
Boosting relies on an iterative estimation procedure to minimize a predefined loss function, where weak learners, often referred to as base-learners, are fitted in every iteration. 
The overall model is obtained by combining these weak learners into an ensemble \citep{Mayr2014}. 
In statistical boosting, the base-learners are statistical models of low complexity, for example simple linear models or splines with a low degree of freedom. 
For fitting (generalized) additive models, each covariate $x_j$ is assigned a base-learner that represents the corresponding effect $f_j, j \in \{1, ..., J\}$. 
The fitted base-learners selected across all iterations $m\in\{1, ..., m_\text{max}\}$, which are typically denoted by $\hat{h}^{[m]}_j, j\in\{1, ..., J\},$ and each only capture part of the effect $f_j$, are then linearly combined to form the effect estimate $\hat{f}_j$. 

This work focuses on component-wise gradient boosting \citep{Buehlmann2003, Buehlmann2007}, which will be referred to as \textit{boosting} from this point on. 
In order to minimize the loss function, in this boosting approach the base-learners are fitted to the negative gradient vector.
In every iteration the model is only updated by the best-fitting base-learner with respect to a least squares criterion and the fitted base-learner is scaled by a small step length factor in order to avoid overfitting.
In general, base-learners can range from simple linear models to trees. 
Here, we focus on regression-type base-learners that are of the form of a penalized linear model. 
The fitted base-learners $\hat{h}_j, j\in\{1, ..., J\}$, applied to the observed covariate values $\boldsymbol{x}_j$, can thus be expressed by 
\begin{align}
\hat{h}_j(\boldsymbol{x}_j) = \boldsymbol{X}_j \left( \boldsymbol{X}_j^\top \boldsymbol{X}_j + \tilde{\lambda}_j \boldsymbol{K}_j \right)^{-1} \boldsymbol{X}_j^\top \boldsymbol{u}, \label{daub:methods_eq_penalized_linear_baselearner}
\end{align}
where $\boldsymbol{u}$ is the negative gradient evaluated at the model in the current iteration, $\boldsymbol{X}_j$ and $\boldsymbol{K}_j$ are the design and penalty matrix of the base-learner associated with $h_j$ and $\tilde{\lambda}_j$ is the respective penalty parameter. 
For example when modeling a linear effect $f_j$, $\boldsymbol{X}_j = \left( \boldsymbol{1}, \boldsymbol{x}_j \right)$ and $\tilde{\lambda}_j=0$. 
An overview of different base-learner types can be found in \cite{Hofner2012}. 

This generic gradient-based boosting algorithm is aimed at estimating a model with a single additive predictor. 
When boosting GAMLSS, the negative log-likelihood
\begin{align*}
-\ell(\theta_1, ..., \theta_K; y)
\end{align*}
is considered as the loss function and $K$ different additive predictors $\eta_{\theta_k}=g_{\theta_k}(\theta_k)$ have to be fitted \citep{Mayr2012, Thomas2018}. 
The gradient-boosting routine outlined above is, therefore, performed for every additive predictor $\eta_{\theta_k}, k \in \{1,...,K\}$. 
In particular, in iteration $m$ the negative gradient 
\begin{align*} 
\boldsymbol{u}^{[m]}_{\theta_k} = \left(- \frac{\partial}{\partial \eta_{\theta_k}} \ell\left(g_{\theta_1}^{-1}(\eta_{\theta_1}), ..., g_{\theta_K}^{-1}(\eta_{\theta_K}); y \right) \middle\vert_{\eta_{\theta_1}=\eta_{\theta_1}^{[m-1]}(\boldsymbol{x}_i), ..., \eta_{\theta_K}=\eta_{\theta_K}^{[m-1]}(\boldsymbol{x}_i), y=y_i} \right)_{i=1, ..., n}
\end{align*}
is computed, all base-learners are fitted to the negative gradient vector and the best fitted base-learner $\hat{h}^{*[m]}_{\theta_k}$ is determined. 
Based on the selected base-learner, the update candidate $\hat{\eta}^{[m-1]}_{\theta_k} + \nu \hat{h}^{*[m]}_{\theta_k}$ is computed, where $\nu$ represents the step length. 
In order to ensure a fair base-learner selection, all base-learners should exhibit a comparable degree of flexibility.
This is typically achieved by assigning the same number of equivalent degrees of freedom to the base-learners \citep{Kneib2009}.

Once all update candidates are determined, the predictor $\hat{\eta}_{\theta_{k^*}}$ that is updated in iteration $m$ is selected by comparing the negative log-likelihood of models with a single update candidate and the remaining predictors set to the values from the previous iteration, i.e.
\begin{align*}
k^* = \underset{k \in \{1, ..., K\}}{\text{arg min}} \,
- \sum_{i=1}^n \ell \left(g_{\theta_1}^{-1}\left(\hat{\eta}^{[m-1]}_{\theta_1}(\boldsymbol{x}_i)\right), \cdots, g_{\theta_k}^{-1}\left(\hat{\eta}^{[m-1]}_{\theta_k}(\boldsymbol{x}_i) + \nu \hat{h}^{*[m]}_{\theta_k}(x_{ij_{\theta_k}^*})\right), \cdots,g_{\theta_K}^{-1}\left(\hat{\eta}^{[m-1]}_{\theta_K}(\boldsymbol{x}_i)\right); y_i \right). 
\end{align*}
The model is then only updated with respect to this best update candidate, which is called \textit{non-cyclical boosting}. 
The non-cyclical boosting algorithm is displayed in Algorithm \ref{daub:algo_non-cyclical_boosting_GAMLSS}, where we on purpose leave the step length $\nu$ more flexible than originally introduced by \cite{Thomas2018}.

\begin{algorithm}[tb]
\caption{Non-cyclical component-wise gradient boosting algorithm for GAMLSS}\label{daub:algo_non-cyclical_boosting_GAMLSS}
\begin{algorithmic}[1]
\State Initialize the predictors $\hat{\eta}_{\theta_1}^{[0]}, ... \, , \hat{\eta}_{\theta_K}^{[0]}$ 
with offset values.
\For{$m=1$ \text{\rm{to}} $m_\text{\rm{stop}}$}
\For{$k=1$ \text{\rm{to}} $K$}
\State Compute the estimated negative gradient vector \\
\hspace*{4em} $\boldsymbol{u}^{[m]}_{\theta_k} = \left(- \frac{\partial}{\partial \eta_{\theta_k}} \ell\left(g_{\theta_1}^{-1}(\eta_{\theta_1}), \cdots, g_{\theta_K}^{-1}(\eta_{\theta_K}); y \right) \middle\vert_{\eta_{\theta_1}=\eta_{\theta_1}^{[m-1]}(\boldsymbol{x}_i), \cdots, \eta_{\theta_K}=\eta_{\theta_K}^{[m-1]}(\boldsymbol{x}_i), y=y_i} \right)_{i=1, ..., n}.$
\State Fit every base-learner to the negative gradient vector: \\
\hspace*{4em} $\left(x_{ij},u_{\theta_k,i}^{[m]}\right)_{i=1, ..., n} \overset{\text{base-learner}}{\longrightarrow} \hat{h}^{[m]}_{j, \theta_k} \text{ for } j = 1, ... \, , J$ 
\State Select the best-fitting base-learner $\hat{h}^{[m]}_{j^*_{\theta_k}, \theta_k}$, where \\
\hspace*{4em}  $j^*_{\theta_k} = \underset{j \in \{1, ..., J\}}{\text{arg min}} \sum_{i=1}^n \left(u_{\theta_k,i}^{[m]} - \hat{h}^{[m]}_{j, \theta_k}(x_{ij})\right)^2$. \label{daub:algo_line_select_base-learner}
\State Choose the step length $\nu_{\theta_k}^{[m]}$.
E.g., $\nu_{\theta_k}^{[m]} = 0.1$. \label{daub:algo_line_steplength}
\State Compute the update candidate \\
\hspace*{4em} $\hat{\eta}_{\theta_k}^{c [m]} = \hat{\eta}_{\theta_k}^{[m-1]} + \nu_{\theta_k}^{[m]} \, \hat{h}^{[m]}_{j^*_{\theta_k}, \theta_k}$ .
\EndFor
\State Select the best update candidate $\hat{\eta}^{c[m]}_{\theta_{k^*}}$, where \\
\hspace*{2.5em} $k^* = \underset{k \in \{1, ... , K\}}{\text{arg min}} - \sum_{i=1}^n \ell \left(g_{\theta_1}^{-1}\left(\hat{\eta}^{[m-1]}_{\theta_1}(\boldsymbol{x}_i)\right), ..., g_{\theta_k}^{-1}\left(\hat{\eta}^{c[m]}_{\theta_k}(\boldsymbol{x}_i)\right), ...,g_{\theta_K}^{-1}\left(\hat{\eta}^{[m-1]}_{\theta_K}(\boldsymbol{x}_i)\right); y \right)$. 
\label{daub:algo_line_select_predictor}
\State Set \\ 
\hspace*{3em} $\hat{\eta}^{[m]}_{\theta_{k}} = 
\begin{cases}
\hat{\eta}_{\theta_k}^{c [m]} &\text{ if } k = k^* \\
\hat{\eta}_{\theta_k}^{[m-1]} &\text{ if } k \neq k^*
\end{cases}$.
\EndFor
\end{algorithmic}
\end{algorithm} 

The iterative update procedure in boosting algorithms is typically stopped before convergence as this yields several advantages.
First, early stopping induces coefficient shrinkage and often improves the predictive performance compared to a fully converged model. 
In addition, if not all covariate effects are selected until the stopping iteration, an intrinsic variable selection is achieved, resulting in a sparser model. 
With the step length typically set to a small, predefined value \citep{Buehlmann2007},
the stopping iteration $m_\text{stop}$ is the main tuning parameter of boosting algorithms.
It is commonly obtained via cross-validation or by other criteria based on out-of-sample predictive performance. 
Alternatively, also information criteria can be applied to find an appropriate stopping iteration \citep{Mayr2012b}.

In non-cyclical boosting algorithms, the relative importance of the additive predictors is determined intrinsically by the sequence of selected predictor updates.
As a result, only a single stopping iteration has to be tuned, like in single-predictor models.
This constitutes a key difference to the \textit{cyclical} updating scheme \citep{Mayr2012}, in which all additive predictors are updated sequentially in each iteration. 
In cyclical boosting, an individual stopping iteration is assigned to each additive predictor, which has the drawback of a more complex and computationally intensive tuning procedure \citep{Hofner2016}.

\subsection{Non-cyclical Boosting with Shrunk Optimal Step Lengths} 

Determining the relative importance of the additive predictors intrinsically has limitations when the negative gradients are on different scales, as
differences in the gradient size $\left\Vert \boldsymbol{u}_{\theta_k} \right\Vert^2$ directly translate to the size of the fitted base-learners $\hat{h}^*_{\theta_k}$ evaluated at $\boldsymbol{x}_{j^*_{\theta_k}}, k \in \{1, ..., K\}$. 
Additive predictors associated with larger gradients therefore have an advantage in the likelihood-based comparison of update candidates due to their larger update size $\left\Vert \nu \hat{h}^*_{\theta_k}(\boldsymbol{x}_{j^*_{\theta_k}})\right\Vert^2$ (Algorithm \ref{daub:algo_non-cyclical_boosting_GAMLSS}, line \ref{daub:algo_line_select_predictor}). 
As a result, these additive predictors tend to be selected more often for update (as long as they have not converged yet), which leads to differences in convergence speed and degree of regularization across predictors at the joint stopping iteration. 
In the final model, the degree of coefficient shrinkage then differs among the additive predictors and the variable selection deteriorates. 

\begin{figure}[!b]
\centering
\begin{subfigure}{.5\textwidth}
  \centering
  \includegraphics[width=\linewidth]{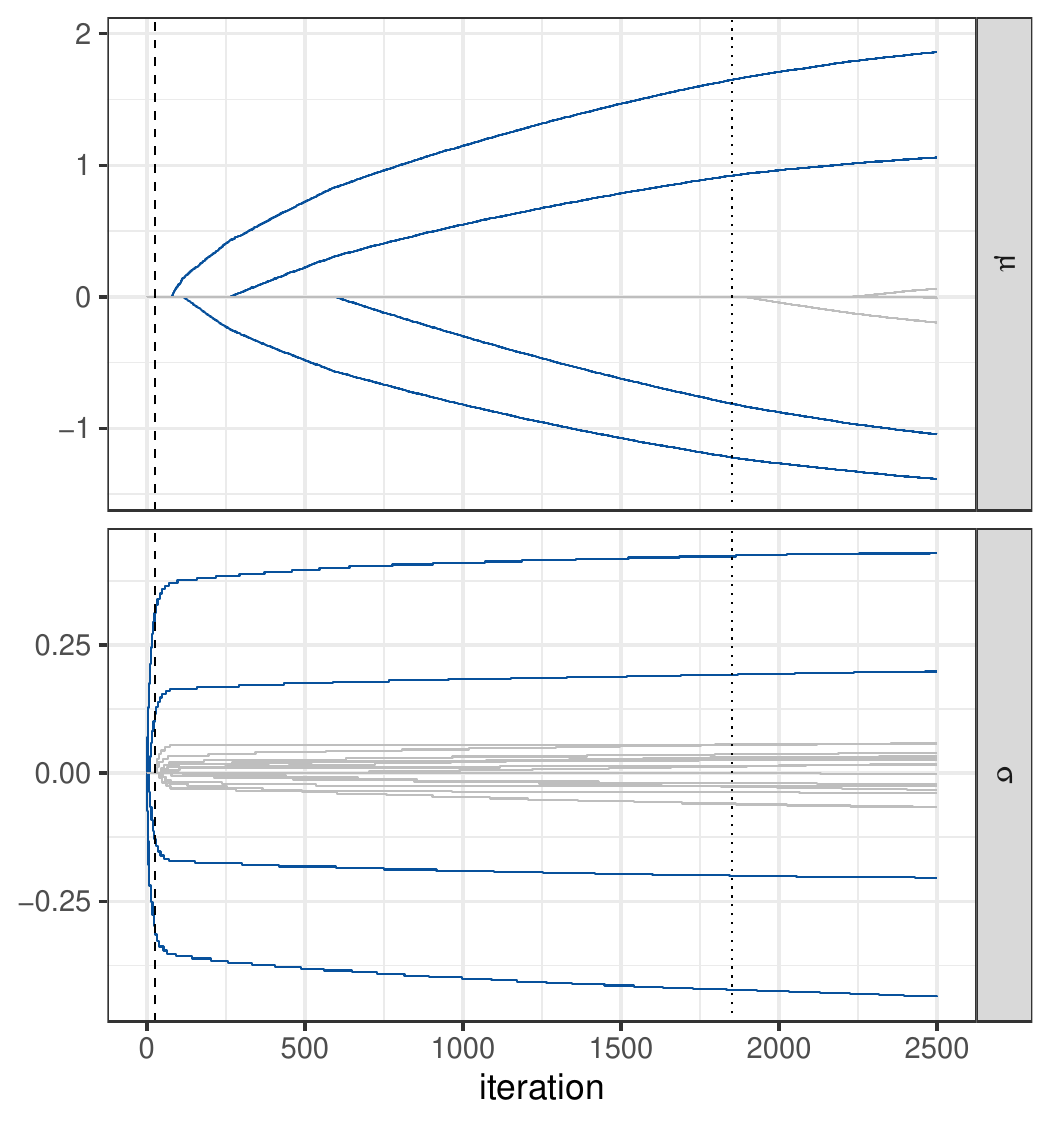}
  \caption{fixed step lengths}
  \label{fig:sub1}
\end{subfigure}%
\begin{subfigure}{.5\textwidth}
  \centering
  \includegraphics[width=\linewidth]{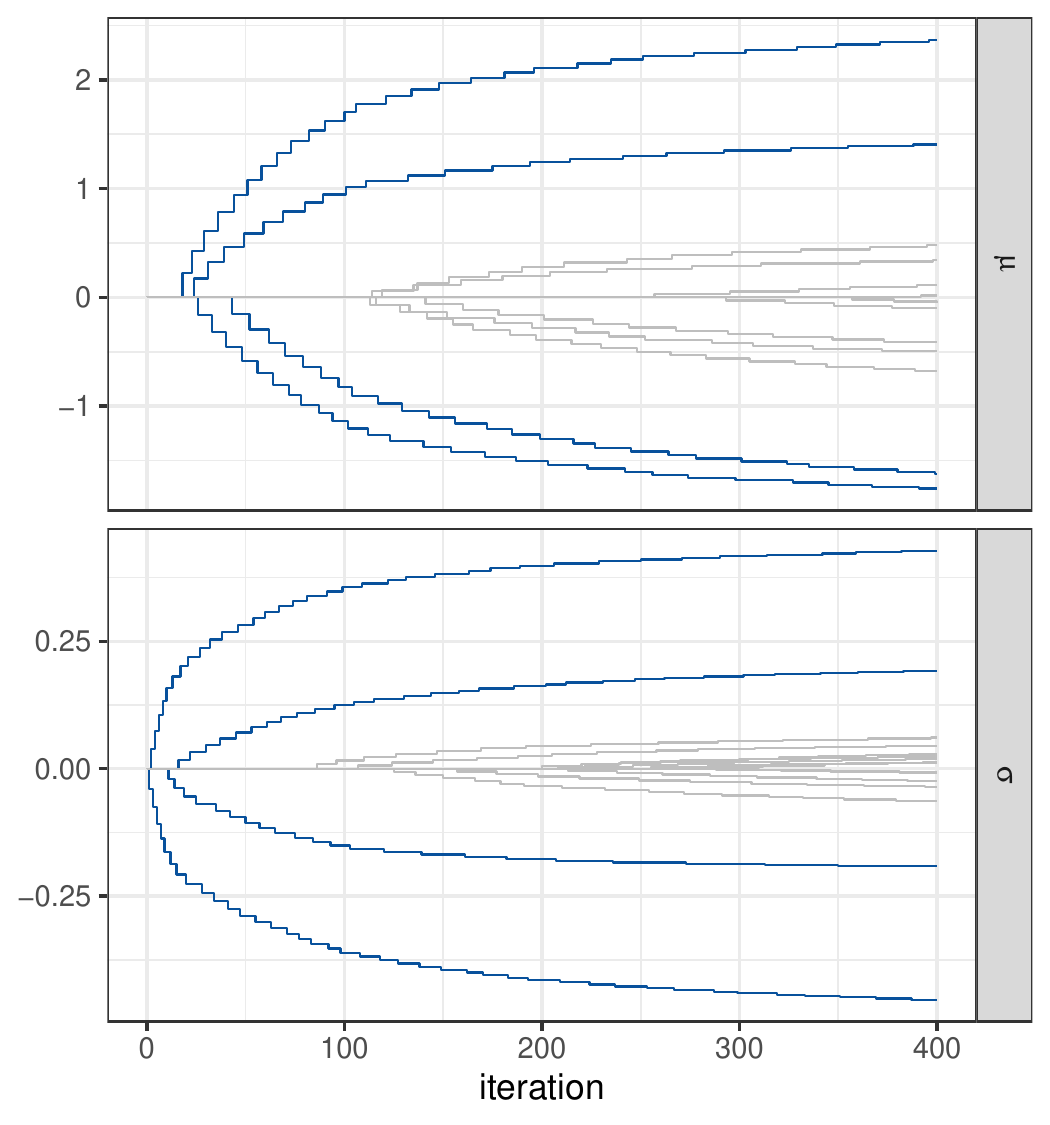}
  \caption{shrunk optimal step lengths}
  \label{fig:sub2}
\end{subfigure}
\caption{Coefficient paths for a Gaussian location and scale model in the simulation setting without additional non-linear effect (\ref{daub:simu_gaussian_setting}). 
Dark blue paths represent informative and gray paths uninformative effects. The dashed and dotted vertical lines represent potential stopping iterations}
\label{daub:fig_gaussian_pathplots}
\end{figure}

Fig.~\ref{daub:fig_gaussian_pathplots}~(a) illustrates this problem. 
Either the more slowly converging additive predictor ($\eta_\mu$ in this example) misses informative covariates and exhibits a strong coefficient shrinkage (dashed exemplary stopping iteration), 
or the faster converging predictor ($\eta_\sigma$ here) includes non-informative covariates and shows almost no shrinkage in the informative effects (dotted exemplary stopping iteration), or a combination of the two (stopping iterations in between).
In the second case, the algorithm needs many iterations until $\eta_\mu$ is adequately fitted, which leads to a long run time. 

A way to address these issues is to replace the fixed step length $\nu$ by a shrunk optimal step length $\nu^{[m]}_{\theta_k}$, which can vary across additive predictors and iterations.
In the context of GAMLSS, this was first introduced by \cite{Zhang2022} for boosting a Gaussian location and scale model with linear effects.
The shrunk optimal step length for distribution parameter $\theta_k$ in iteration $m$ is defined as 
{
\begin{align}
&\nu^{[m]}_{\theta_k} = \lambda \nu^{*[m]}_{\theta_k} \, \text{ with } \nonumber \\
&\nu_{\theta_k}^{*[m]} = \underset{\nu}{\text{arg min}} \,
- \sum_{i=1}^n \ell \left(g_{\theta_1}^{-1}\left(\hat{\eta}^{[m-1]}_{\theta_1}(\boldsymbol{x}_i)\right), \cdots, g_{\theta_k}^{-1}\left(\hat{\eta}^{[m-1]}_{\theta_k}(\boldsymbol{x}_i) + \nu \hat{h}^{*[m]}_{\theta_k}(x_{ij_{\theta_k}^*})\right), \cdots,g_{\theta_K}^{-1}\left(\hat{\eta}^{[m-1]}_{\theta_K}(\boldsymbol{x}_i)\right); y_i \right) ,
\label{daub:eq_opt_sl}
\end{align}
}%
where $\lambda$ represents a shrinkage factor, e.g. $\lambda=0.1$.
When using a shrunk optimal step length, the size of the candidate update $\nu_{\theta_k} \hat{h}^*_{\theta_k}$ evaluated at $\boldsymbol{x}_{j^*_{\theta_k}}$ reflects its potential to reduce the loss function.
By that, structurally small fitted base-learners can be compensated and the size of candidate updates corresponding to a small negative gradient is increased.
This has two implications. First, the candidate updates are of similar size and thus a fairer decision of which additive predictor is updated can be achieved. 
Second, given a candidate update corresponding to small negative gradients is selected the updates are larger and fewer updates are required until convergence. 
This leads to a larger relative convergence speed of additive predictors with a small negative gradient and thus a more similar convergence speed among predictors.
Due to the more balanced convergence behavior, using shrunk optimal step lengths results in a similar degree of regularization among additive predictors and an improved variable selection in the overall model (see Fig.~\ref{daub:fig_gaussian_pathplots}~(b)).

Increasing the convergence speed of originally slowly converging additive predictors, moreover, reduces the number of iterations required until the stopping criterion is reached and therefore the overall run time. 
The absolute convergence speed can be adjusted by modifying the shrinkage factor $\lambda$. 
The shrinkage factor underlies the usual trade-off between size of the step length and stopping iteration also observed for fixed step lengths  \citep{Friedman2001,Buehlmann2007}.

\subsection{Combining Shrunk Optimal Step Lengths with Non-linear Base-learners}
\label{daub:subsection_opt_sl_base-learners}

\begin{figure}[!b]
\includegraphics[width=\textwidth]{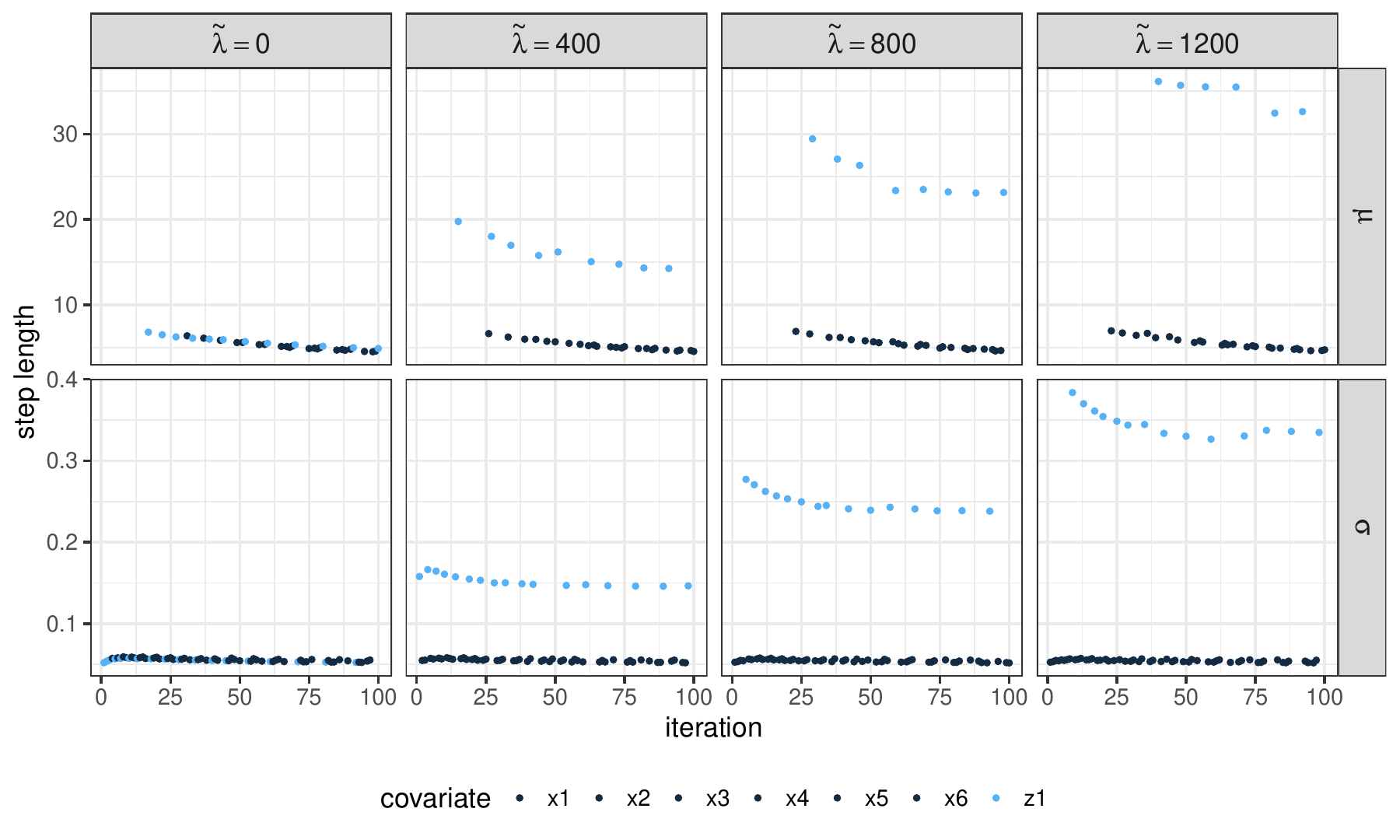}
\caption{Shrunk optimal step lengths for varying levels of penalization of the base-learner representing the categorical effect (columns) in the Gaussian simulation setting (\ref{daub:simu_gaussian_setting})
\label{daub:methods_gaussian_steplengths_lambda}}
\end{figure} 

\begin{figure}[!tb]
\includegraphics[width=\textwidth]{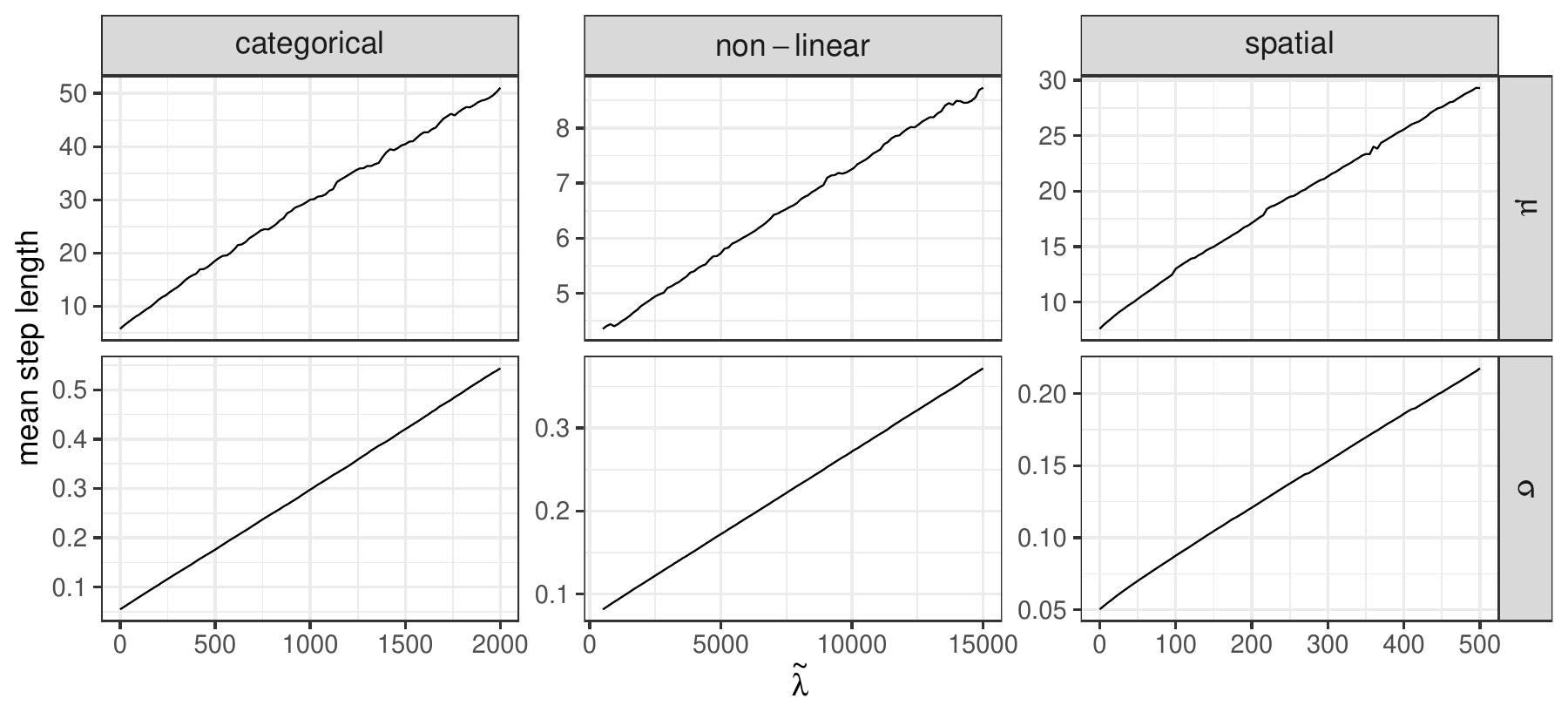}
\caption{Comparison of the penalty parameter and the mean step length in the first 100 iterations of a simulation run for base-learners representing different effects (columns) in the Gaussian simulation setting (\ref{daub:simu_gaussian_setting}). For the explicit specification of the base-learners, see Sect.~\ref{daub:section_simulations} \label{daub:methods_gaussian_lambda_sl}}
\end{figure}

Until now, shrunk optimal step lengths have only been combined with simple linear base-learners. 
In principle, the outlined mechanism of compensating for small negative gradients by increasing the update size of a fitted base-learner works in the same way also for other base-learner types.
There is, however, one important difference when considering penalized base-learners like splines, Markov random fields or jointly dummy-coded categorical effects with reduced equivalent degrees of freedom. 
If the coefficient size is regularized by a base-learner, its potential for improving the loss function in the direction of the fit $\hat{h}_j(\boldsymbol{x}_{j})$ is not exhausted. 
This is picked up by the optimal step length leading to a larger step length than for unpenalized base-learners of the same additive predictor.
The magnitude of the difference between the optimal step lengths depends on the strength of the penalty. 
Fig.~\ref{daub:methods_gaussian_steplengths_lambda} illustrates this association for a jointly dummy-coded effect with five categories and varying levels of penalization. 
More specifically, the relationship between the optimal step length level and the penalty parameter $\tilde{\lambda}$ of a penalized linear base-learner is approximately linear, as illustrated in Fig.~\ref{daub:methods_gaussian_lambda_sl} for a Gaussian location and scale model and different base-learner types. 
This relationship, however, is only as distinct when considering the same base-learner.
Across base-learner types and covariates, the optimal step length is influenced by multiple additional factors beyond the penalty parameter.

With the equivalent degrees of freedom of a penalized linear base-learner associated with $h_j$ defined as \citep{Hastie1990} 
\begin{align*}
\text{df}_j (\tilde{\lambda}_j) = \text{trace}(\boldsymbol{X}_j \left( \boldsymbol{X}_j^\top \boldsymbol{X}_j + \tilde{\lambda}_j \boldsymbol{K}_j \right)^{-1} \boldsymbol{X}_j^\top) ,
\end{align*}
the relationship between the number of equivalent degrees of freedom and the corresponding optimal step length cannot be isolated as cleanly and depends on the concrete base-learner specification and the data.
For example when reducing the number of equivalent degrees of freedom of a categorical effect,  
the penalty parameter and thus the optimal step length is comparatively large as the penalty matrix $\boldsymbol{K}$ is an identity matrix exclusively penalizing the size of the coefficients. 
A Markov random field base-learner with the same number of regions would have a lower $ \tilde{\lambda}$ and thus a lower optimal step length as not solely the size but also the variation of effects of neighboring regions is penalized. 

The positive relationship between degree of penalization and optimal step length constitutes an additional advantage of shrunk optimal step lengths when penalized base-learners are included in the model. 
With fixed step lengths, penalized base-learners require more iterations until the effect is sufficiently fitted. 
Under strong penalization, this leads to a long run time and may even affect the variable selection. 
Reducing the penalization mitigates these issues, but simultaneously gives the base-learner an advantage over less flexible base-learners. 
As shrunk optimal step lengths increase the update size after the base-learner selection, the number of iterations needed until the base-learner is sufficiently fitted is reduced, 
while a fair base-learner selection can be maintained.

\subsection{Computing Shrunk Optimal Step Lengths for ZINB-GAMLSS} 

Optimal step lengths can be determined in different ways.
Most commonly, optimal step lengths are obtained numerically via line search as proposed by \cite{Friedman2001} and \cite{Zhang2022}. 
Using a line search, however, has two notable drawbacks.
First, an appropriate search interval must be specified that contains the optimum while not being unnecessarily large, 
as overly wide intervals increase the computational cost and may lead to numeric instabilities. 
Choosing a suitable upper bound can be challenging, in particular when optimal step lengths are large for certain distribution parameters or when strongly penalized base-learners are included in the boosting specification.
Second, performing a line search to determine the optimal step length for each update candidate in every iteration substantially increases the computation time, which is especially pronounced for complex loss functions and large search intervals.
As an alternative, analytical expressions or analytical approximations for optimal step lengths can be derived, which has been done for some response variable distributions and distribution parameters \citep{Zhang2022, Daub2025}. 
However, as the complexity of the distribution increases, the expressions of the approximated shrunk optimal step lengths get very lengthy and the accuracy of the approximations degrades.
Deriving analytical approximations of the optimal step lengths is, therefore, not suitable here.

To avoid the outlined issues associated with a line search and analytical approximations, 
we determine the optimal step lengths using Newton's method applied to the first-order condition of the optimization problem in (\ref{daub:eq_opt_sl}).
The method is implemented using the R package \textit{nleqslv} \citep{nleqslv2023} and yields very similar results as a line search while being computationally more efficient.
Since optimal step lengths for a given base-learner and additive predictor tend to be similar across boosting iterations, 
the Newton algorithm is initialized using the optimal step length from the iteration in which the base-learner was selected last.
If a base-learner has not yet been updated, the initial value is set to one. 
The first-order conditions are derived from the negative log-likelihood of the ZINB-GAMLSS given by
\begin{align*}
-\sum_{i=1}^n &\ell\left(\mu_i, \alpha_i, \pi_i; y_i \right) \\
= &-\sum_{i: y_i=0}  \ln\left(\pi_i + (1-\pi_i)(1+\alpha_i \mu_i)^{-\frac{1}{\alpha_i}}\right) \\
\hspace*{1em}&- \sum_{i: y_i>0} \ln(1-\pi_i) + y_i \ln \left( \frac{\alpha_i \mu_i}{1+\alpha_i \mu_i}\right) - \frac{1}{\alpha_i} \ln(1+\alpha_i \mu_i) + \ln\Gamma\left(y_i+\frac{1}{\alpha_i}\right) - \ln\Gamma(y_i+1) - \ln\Gamma\left(\frac{1}{\alpha_i}\right),
\end{align*}
where $\ln\Gamma$ denotes the log-gamma function. 
For the optimal step length of $\mu$ in iteration $m$ we for example obtain
\begin{align}
0 &= - \frac{\partial}{\partial \nu_\mu} \sum_{i=1}^n \ell\left(\exp\left(\hat{\eta}_{\mu,i}^{[m-1]} + \nu_\mu \hat{h}_{\mu,i}^{*[m]}\right), \hat{\alpha}^{[m-1]}_i, \hat{\pi}^{[m-1]}_i; y_i \right) \Bigg\vert_{\nu_\mu=\nu_\mu^{*[m]}} \nonumber \\
&= \sum_{i:y_i=0} \frac{\exp\left(\hat{\eta}_{\mu,i}^{[m-1]} + \nu_\mu^{*[m]} \hat{h}_{\mu,i}^{*[m]}\right) \hat{h}_{\mu,i}^{*[m]}}
{\frac{\hat{\pi}^{[m-1]}_i}{1-\hat{\pi}^{[m-1]}_i} \left(1 + \hat{\alpha}^{[m-1]}_i \exp\left(\hat{\eta}_{\mu,i}^{[m-1]} + \nu_\mu^{*[m]} \hat{h}_{\mu,i}^{*[m]}\right)\right)^{\frac{1}{\hat{\alpha}^{[m-1]}_i}+1} + \left(1 + \hat{\alpha}^{[m-1]}_i \exp\left(\hat{\eta}_{\mu,i}^{[m-1]} + \nu_\mu^{*[m]} \hat{h}_{\mu,i}^{*[m]}\right) \right)} \nonumber \\
&\hspace{2em} - \sum_{i:y_i>0} \frac{\hat{h}_{\mu,i}^{*[m]} \left(y_i - \exp\left(\hat{\eta}_{\mu,i}^{[m-1]} + \nu_\mu^{*[m]} \hat{h}_{\mu,i}^{*[m]}\right)\right)}{1 + \hat{\alpha}_i^{[m-1]} \exp\left(\hat{\eta}_{\mu,i}^{[m-1]} + \nu_\mu^{*[m]} \hat{h}_{\mu,i}^{*[m]}\right)} ,
\label{daub:eq_FOC_nu_mu}
\end{align}
where for notational convenience we define $\hat{\eta}_{\theta_k,i}^{[m-1]} = \hat{\eta}_{\theta_k}^{[m-1]}(\boldsymbol{x}_i)$, $\hat{h}_{\theta_k,i}^{*[m]} = \hat{h}_{\theta_k}^{*[m]}(\boldsymbol{x}_i)$ and $\hat{\theta}_{k,i}^{[m]} = \hat{\theta}_k^{[m]}(\boldsymbol{x}_i)$.
The first-order conditions for $\nu_\alpha^{*[m]}$ and $\nu_\pi^{*[m]}$ can be derived analogously. 
Further details on the derivation of equation~(\ref{daub:eq_FOC_nu_mu}) as well as the remaining first-order conditions are provided in the Supplement.

\section{Simulation Study}
\label{daub:section_simulations}

To evaluate the performance of the non-cyclical boosting algorithm with shrunk optimal step lengths, we conducted a simulation study with two response variable distributions and different effect types, which are motivated by our application. 
In a first Gaussian setting, we investigate combining shrunk optimal step lengths with other than simple linear base-learners,
focusing on the selection and degree of shrinkage of the effects.
Secondly, we consider a zero-inflated negative binomial response variable in order to investigate if shrunk optimal step lengths yield a more balanced regularization and variable selection behavior among additive predictors also for this model. 
In both settings, the results for shrunk optimal step lengths are compared to those from the fixed step length approach, using a fixed step length and shrinkage factor of 0.1.
To determine the stopping iteration, we apply a robust 10-fold cross-validation variant that selects the earliest iteration within a 2\% range of the minimum, as this yields more stable results in the presence of a flat likelihood. 
In each simulation setting, 100 runs are conducted.
Estimation via boosting is carried out with our extension of the R add-on package \textit{gamboostLSS} \citep{Hofner2016}.
For comparison, the models are also estimated by maximum likelihood using the R add-on package \textit{gamlss} \citep{Stasinopoulos2008}.

\subsection{Gaussian setting}
\label{daub:subsection_simu_gaussian}

In the first setting, we generate $n=500$ observations $y_i$ from $Y_i \sim \mathcal{N}(\mu(\boldsymbol{x}_i), \sigma(\boldsymbol{x}_i))$ with 
\begin{align}
\mu(\boldsymbol{x}_i) &= \eta_\mu(\boldsymbol{x}_i) =  -1.5 x_{1i} + 2.5 x_{2i} + 1.5 x_{3i} - 2.5 x_{4i} + f_\mu(z_i) \nonumber \\ 
\log(\sigma(\boldsymbol{x}_i)) &= \eta_\sigma(\boldsymbol{x}_i) = 2 + 0.2 x_{3i} + 0.5 x_{4i} - 0.2 x_{5i} - 0.5 x_{5i} + f_\sigma(z_i). 
\label{daub:simu_gaussian_setting}
\end{align}
The covariates $x_{1i}, x_{3i}, x_{5i}$ are independently drawn from $\mathcal{U}(-1,1)$ and $x_{2i}, x_{4i}, x_{6i}$ are independent realizations of $\mathcal{B}(1, 0.5)$, $i \in \{1, ..., n\}$.
In addition, 20 non-informative covariates are included, where we have $X_{7i}, X_{9i}, ..., X_{25i} \sim \mathcal{U}(-1,1)$ and $X_{8i}, X_{10i}, ..., X_{26i} \sim \mathcal{B}(1, 0.5)$.
The function $f$ represents different additional effects: either a categorical effect, a non-linear effect or a discrete spatial effect. These are considered as different settings;
details on the specification of these effects are provided in the Supplement.
When an informative categorical or non-linear effect is included, a non-informative effect is added as well. 
For the discrete spatial effect, the informative and non-informative effect are considered separately.

Boosting is performed using linear base-learners with intercept for covariates
$X_1, ..., X_{26}$, a linear base-learner with joint dummy-coded categories for the categorical effect, a centered P-spline of degree three with a second order difference penalty and 20 inner knots for the non-linear effect, and a centered Markov random field base-learner with the corresponding neighborhood structure for the discrete spatial effect.
To ensure a fair base-learner selection, the effective degrees of freedom are set to two for all base-learners \citep{Hofner2011}.
As this setting is designed to show the imbalancedness problem outlined in the previous section, the data exhibit high variability and one of the gradients ($\boldsymbol{u}_\mu$) is very small.
Consequently, the model is generally difficult to estimate, and the variability in the estimated effects is comparatively large for all methods. 
Note that in balanced settings,  fixed and shrunk optimal step lengths yield very similar results \citep{Zhang2022}.

\begin{figure}[!b]
\includegraphics[width=\textwidth]{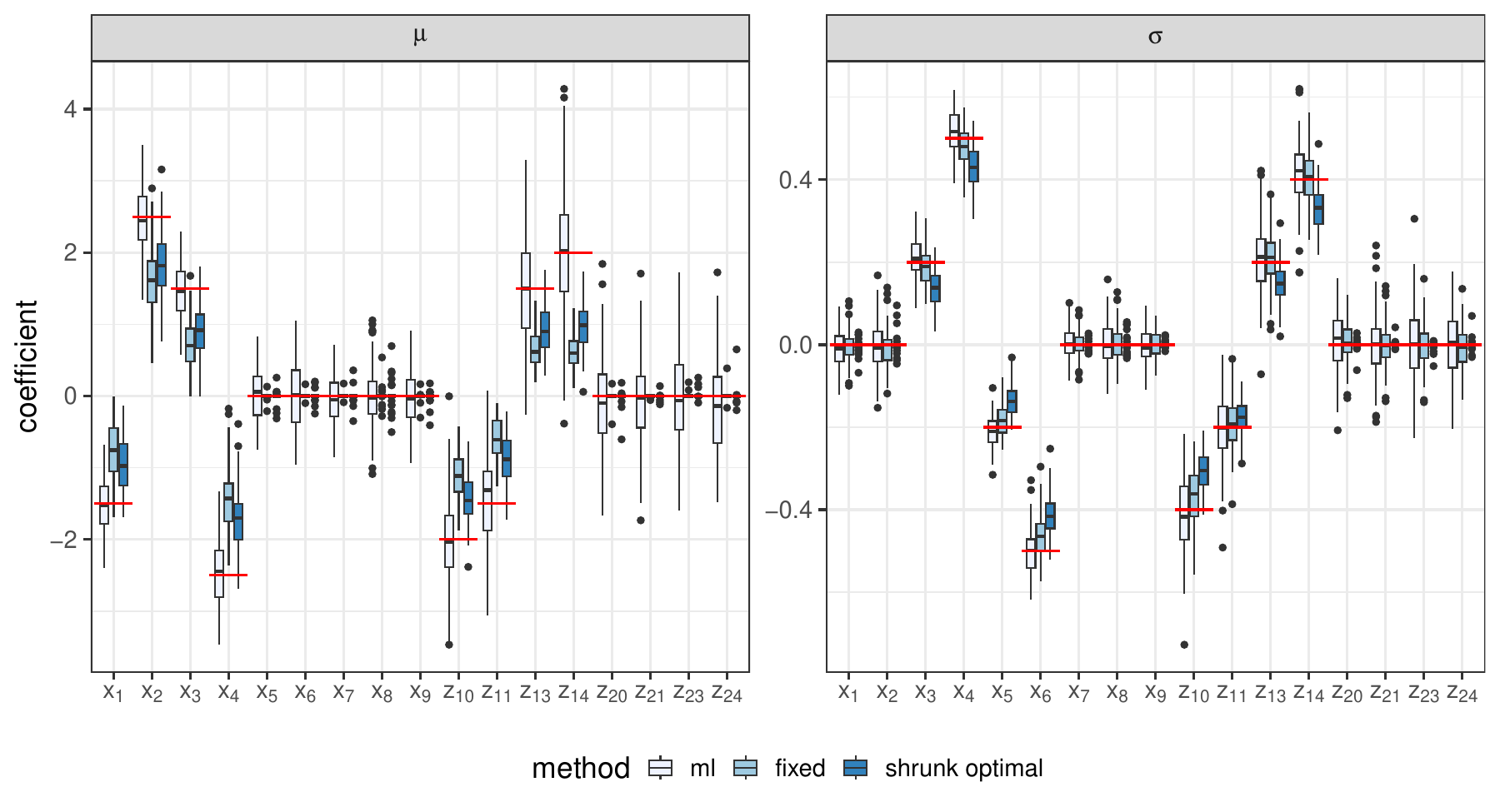}
\caption{Distribution of the coefficient estimates in the Gaussian simulation setting (\ref{daub:simu_gaussian_setting}) with categorical effects. The red horizontal lines represent the true coefficients \label{daub:simu_gaussian_coefficient_boxplots}}
\end{figure}

Fig. \ref{daub:simu_gaussian_coefficient_boxplots} shows the distribution of the coefficient estimates in the setting with categorical effects. 
Note that shrinkage of the estimated effects is typically intended when applying boosting and can be reduced by stopping the algorithm later.
With respect to the relative regularization across additive predictors, we find that the fixed step length approach results in relatively weak regularization for $\sigma$ and considerably stronger regularization for $\mu$.
The use of shrunk optimal step lengths increases the regularization of $\sigma$ and reduces the regularization of $\mu$, leading to a considerably smaller imbalance in the degree of regularization between the two additive predictors. 
These finding are in line with the results of \cite{Zhang2022}. 
Comparing categorical with simple linear effects, we find that they behave similarly in terms of relative regularization across additive predictors, with the categorical effect being slightly less regularized overall.

Along with the weaker coefficient shrinkage, the fixed step length approach includes substantially more non-informative covariates in $\sigma$ compared to the shrunk optimal step length approach. 
This is particularly pronounced for the categorical effect, where the non-informative categorical effect is selected in every simulation run (see Table~\ref{daub:table_coefficient_estimates}, columns~1-4). 
For $\mu$, the differences in the selection behavior between the two step length approaches are less pronounced: 
the fixed step length approach includes slightly fewer non-informative effects but does not select the informative effects in all runs. 

\begin{table}[!t]
\caption{Number of selections of the respective covariate effect in the Gaussian simulation setting (\ref{daub:simu_gaussian_setting}) with additional categorical (columns 1-4), non-linear (columns 5-8) and a non-informative spatial effect (columns 9-12) for 100 simulation runs. The informative effects are marked in bold
\label{daub:table_coefficient_estimates}}
\tabcolsep=0pt
\begin{tabular*}{\textwidth}{@{\extracolsep{\fill}}lcccccccccccc@{\extracolsep{\fill}}}
\toprule%
& \multicolumn{4}{@{}c@{}}{categorical} & \multicolumn{4}{@{}c@{}}{non-linear} & \multicolumn{4}{@{}c@{}}{spatial}\\ 
\cmidrule{2-5}\cmidrule{6-9}\cmidrule{10-13}%
& \multicolumn{2}{@{}c@{}}{$\eta_\mu$} & \multicolumn{2}{@{}c@{}}{$\eta_\sigma$} & \multicolumn{2}{@{}c@{}}{$\eta_\mu$} & \multicolumn{2}{@{}c@{}}{$\eta_\sigma$} & \multicolumn{2}{@{}c@{}}{$\eta_\mu$} & \multicolumn{2}{@{}c@{}}{$\eta_\sigma$}\\
\cmidrule{2-3}\cmidrule{4-5}\cmidrule{6-7}\cmidrule{8-9}\cmidrule{10-11}\cmidrule{12-13}%
& fixed & opt & fixed & opt & fixed & opt & fixed & opt & fixed & opt & fixed & opt \\ 
\midrule
$x_1$ & \textbf{98} & \textbf{100} & 81 & 16 & \textbf{100} & \textbf{100} & 70 & 8 & \textbf{86} & \textbf{99} & 69 & 8 \\ 
  $x_2$ & \textbf{100} & \textbf{100} & 81 & 13 & \textbf{100} & \textbf{100} & 75 & 15 & \textbf{99} & \textbf{100} & 75 & 7 \\ 
  $x_3$ & \textbf{94} & \textbf{99} & \textbf{100} & \textbf{100} & \textbf{100} & \textbf{100} & \textbf{100} & \textbf{100} & \textbf{82} & \textbf{93} & \textbf{100} & \textbf{100} \\ 
  $x_4$ & \textbf{100} & \textbf{100} & \textbf{100} & \textbf{100} & \textbf{100} & \textbf{100} & \textbf{100} & \textbf{100} & \textbf{96} & \textbf{100} & \textbf{100} & \textbf{100} \\ 
  $x_5$ & 3 & 7 & \textbf{100} & \textbf{100} & 0 & 8 & \textbf{100} & \textbf{100} & 1 & 5 & \textbf{100} & \textbf{100} \\ 
  $x_6$ & 2 & 8 & \textbf{100} & \textbf{100} & 1 & 7 & \textbf{100} & \textbf{100} & 0 & 1 & \textbf{100} & \textbf{100} \\ 
  $x_7$ & 2 & 7 & 84 & 10 & 4 & 10 & 69 & 19 & 1 & 7 & 64 & 8 \\ 
  $x_8$ & 9 & 14 & 80 & 15 & 4 & 12 & 65 & 9 & 1 & 2 & 57 & 3 \\ 
  $x_9$ & 4 & 9 & 86 & 12 & 1 & 7 & 54 & 9 & 0 & 2 & 70 & 11 \\ 
  $x_{10}$ & 3 & 12 & 80 & 16 & 2 & 8 & 65 & 16 & 0 & 3 & 74 & 10 \\ 
  $z_1$ & \textbf{100} & \textbf{100} & \textbf{100} & \textbf{100} & \textbf{100} & \textbf{100} & \textbf{100} & \textbf{100} & \textbf{--} & \textbf{--} & \textbf{--} & \textbf{--} \\ 
  $z_2$ & 2 & 5 & 100 & 10 & 5 & 21 & 99 & 32 & 0 & 11 & 99 & 11 \\
\bottomrule
\end{tabular*}
\end{table}

\begin{figure}[!b]
\includegraphics[width=\textwidth]{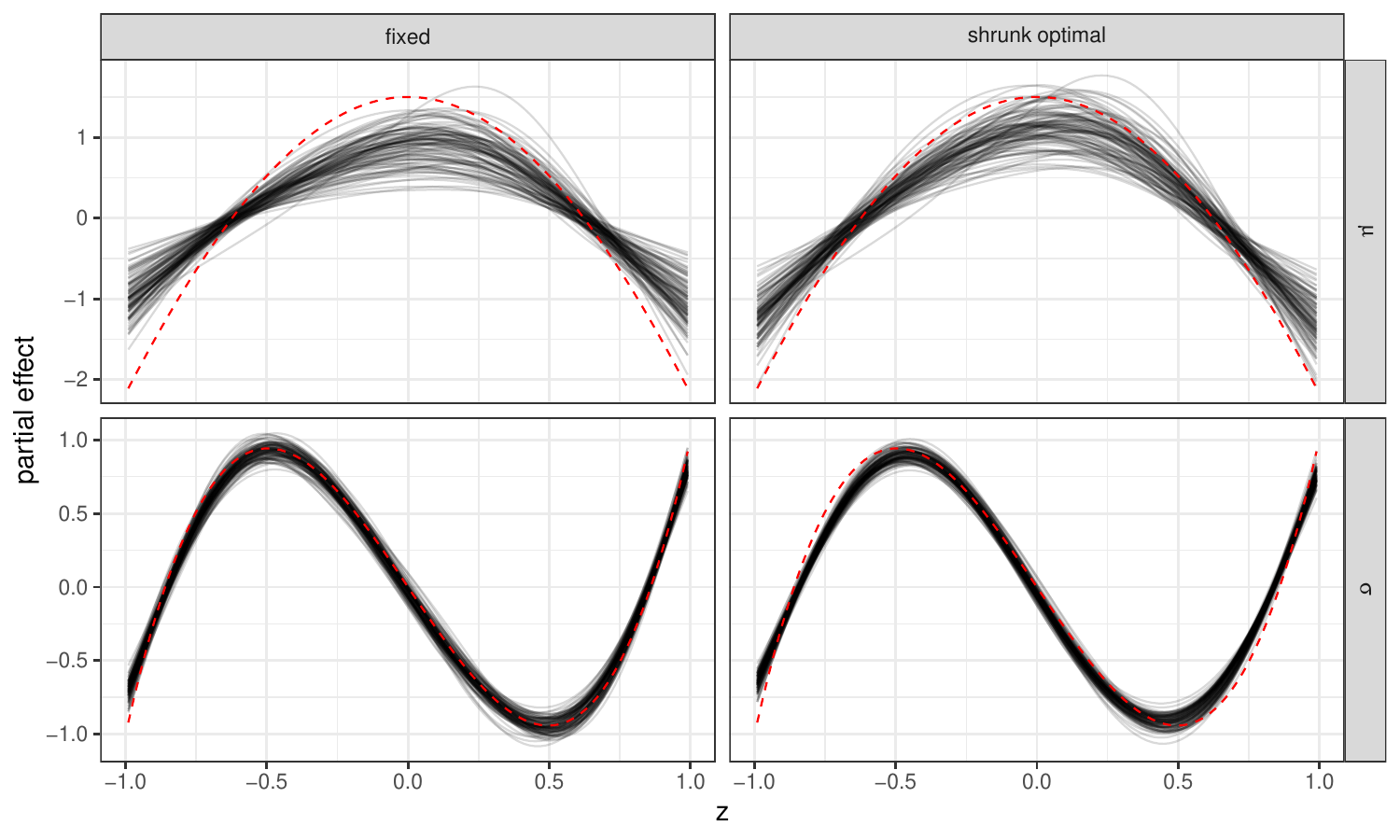}
\caption{Partial effects of the informative non-linear effect in the Gaussian simulation setting (\ref{daub:simu_gaussian_setting}). The red dashed lines represent the true partial effect
\label{daub:simu_gaussian_partial_inf}}
\end{figure}

In the simulation setting with non-linear effects, a similar pattern with respect to the relative regularization of $\mu$ compared to $\sigma$ can be observed, where the non-informative non-linear effect is selected substantially less frequently in the optimal step length approach than when fixed step lengths are used. 
Unlike the categorical effect, the non-informative spline is selected more often in both methods than the linear non-informative effects (see Table~\ref{daub:table_coefficient_estimates}, columns~5-8). 
This likely results from the higher flexibility of the spline base-learner, despite having the same effective degrees of freedom \citep{Hofner2011}.
With respect to the shrinkage of the informative non-linear effects, 
the difference between the step length approaches is small.
The effect for $\mu$ is slightly less shrunk for fixed step lengths and the variability in the effect for $\sigma$ is slightly larger compared to the shrunk optimal step length approach (see Fig.~\ref{daub:simu_gaussian_partial_inf}).

Considering an informative spatial effect yields similar results as the categorical setting. 
Since the region-specific effects are smoothed across neighboring regions, they are expected to exhibit differing levels of shrinkage, 
and consequently also the maximum likelihood estimates are regularized. 
In addition, slightly fewer covariates are selected overall in both spatial settings compared to the categorical setting, which is likely due to the missing second effect in this setting. 
Further results can be found in the Supplement.

In order to evaluate the predictive performance, 
we consider the mean continuous ranked probability score (CRPS) with either the true distribution or the true point value as reference, the negative log-likelihood and the mean squared error (MSE) on test data. 
With respect to both mean CRPS and the negative log-likelihood, 
shrunk optimal step lengths yield a better predictive performance than fixed step lengths in all settings, where the strongest improvement is observed for the CRPS relative to the true distribution. 
The distributions of predictive performance measures are displayed in the Supplement. 

Comparing further properties of the algorithm, we find that using shrunk optimal step lengths reduces the number of iterations needed until stopping substantially, which is in line with previous findings in this regard \citep{Zhang2022, Daub2025}.
In addition, considering more strongly penalized base-learners does not increase the stopping iteration substantially when using shrunk optimal step lengths while the fixed step length approach stops noticeably later (see Table~\ref{daub:table_mstop_gaussian}). 
The difference between fixed and shrunk optimal approach is especially pronounced for the categorical effect, which is due to its penalty structure. 
This supports the discussion in section \ref{daub:subsection_opt_sl_base-learners} and constitutes an enhanced advantage of using shrunk optimal step lengths in combination with penalized base-learners. 

In addition to examining categorical, non-linear and spatial effects separately, we also consider a setting in which all effects are included simultaneously.
The results are consistent with those obtained in the separate settings, with the required number of iterations increasing approximately proportionally, thereby amplifying the difference in stopping iteration.

\begin{table}[!h]
\caption{Mean and standard deviation (in brackets) of the stopping iterations for fixed (top rows) and shrunk optimal step lengths (bottom rows) in the different Gaussian simulation settings. 
In addition, the relative comparison of effective degrees of freedom of two and four (\% mean) is displayed
\label{daub:table_mstop_gaussian}}
\tabcolsep=0pt
\begin{tabular*}{\textwidth}{@{\extracolsep{\fill}}lcccc@{\extracolsep{\fill}}}
\toprule%
& & categorical & non-linear & spatial\\ 
\midrule
 & $\text{df} = 2$ & 3,294 (663) & 1,857 (242) & 2,821 (496)\\ 
fixed & $\text{df} = 4$ & 1,841 (443) & 1,545 (219) & 2,248 (407)\\ 
\cmidrule{2-5}%
 & \% mean & 1.789 & 1.202 & 1.255\\ 
  \midrule
 & $\text{df} = 2$ & 152 (22) & 187 (25) & 139 (21)\\ 
shrunk optimal & $\text{df} = 4$ & 146 (21) & 175 (21) & 136 (20)\\ 
\cmidrule{2-5}%
 & \% mean & 1.041 & 1.069 & 1.022\\ \bottomrule
\end{tabular*}
\end{table}

\subsection{ZINB setting}

In the second setting, $n=4{,}000$ observations are generated from $Y_i \sim \text{ZINB}(\mu(\boldsymbol{x}_i), \alpha(\boldsymbol{x}_i), \pi(\boldsymbol{x}_i))$ with
\begin{align}
\log(\mu(\boldsymbol{x}_i)) &= \eta_\mu(\boldsymbol{x}_i) =  1.8 + 0.2 x_{1i} - 0.35 x_{2i} -0.2 x_{3i} + 0.35 x_{4i} + f_\mu(z_i) \nonumber \\
\log(\alpha(\boldsymbol{x}_i)) &= \eta_\alpha(\boldsymbol{x}_i) = -1.1 + 0.6 x_{2i} + 0.5 x_{3i} - 0.6 x_{4i} - 0.5 x_{5i} + f_\alpha(z_i)  \nonumber \\ 
\text{logit}(\pi(\boldsymbol{x}_i)) &= \eta_\pi(\boldsymbol{x}_i) = -0.8 + x_{3i} - 1.25 x_{4i} - x_{5i} + 1.25 x_{6i} + f_\pi(z_i). 
\label{daub:simu_ZINB_setting}
\end{align}
Again, $x_{1i}, x_{3i}, ..., x_{25i}$ are drawn independently from $\mathcal{U}(-1,1)$, $x_{2i}, x_{4i}, ..., x_{26i}$ are independent realizations of $\mathcal{B}(1, 0.5)$ and categorical, non-linear or spatial effects are added (for more details on the specification, see the Supplement).
The base-learners are specified as in the Gaussian setting. 
Due to the construction of the distribution and the dependence among the distribution parameters, a zero-inflated negative binomial model for location, scale and shape is considerably more difficult to estimate than a Gaussian location and scale model. 
The small gradient of $\alpha$ further amplifies the difficulty.

Overall, our findings for the zero-inflated negative binomial model are consistent with the results from the Gaussian setting. 
However, a larger variability in the coefficient estimates for the maximum likelihood reference was observed and the additive predictor for $\alpha$ is quite strongly regularized in both boosting approaches. 
Like in the Gaussian setting, shrunk optimal step lengths yield a more similar degree of regularization among the additive predictors: 
$\alpha$ is less regularized, while $\mu$ and $\pi$ show a stronger regularization compared to the fixed step length approach. 
These differences are, however, less pronounced than in the Gaussian case, which can be attributed to the ZINB setting being slightly less imbalanced. 

In terms of variable selection, the shrunk optimal approach clearly outperforms the fixed step length approach (see Table~\ref{daub:table_zinb_covsel}).
Compared to fixed step lengths, it substantially reduces the number of false positives in $\mu$ and false negatives in $\alpha$, although it still tends to select more non-informative effects for $\mu$ than for the other additive predictors.
For $\pi$, all informative effects and only few non-informative effects are selected, with $x_2$ being the only exception. 
These selection difficulties likely arise from the challenge of disentangling effects across the dependent additive predictors, and the use of shrunk optimal step lengths does not substantially improve the number of false positives in this case.
Further simulation results are provided in the Supplement. 

Comparing the methods with respect to the predictive performance, we find that the boosting approaches outperform the maximum likelihood reference in terms of CRPS and maximum likelihood. 
Between the boosting approaches, no clear improvement of the shrunk optimal step length approach could be observed. 
Given the lower number of selected covariates, the predictive performance is, however, achieved with a sparser model for shrunk optimal step lengths.
The distributions of the predictive performance measures for the different approaches are displayed in the Supplement. 

\begin{table}[!t]
\caption{Number of selections of the covariate effects in the ZINB simulation setting (\ref{daub:simu_ZINB_setting}) with additional categorical (columns 1-6) and non-linear effects (columns 7-12)  for 100 simulation runs. The informative effects are marked in bold \label{daub:table_zinb_covsel}}
\tabcolsep=0pt
\begin{tabular*}{\textwidth}{@{\extracolsep{\fill}}lcccccccccccc@{\extracolsep{\fill}}}
\toprule%
& \multicolumn{6}{@{}c@{}}{categorical} & \multicolumn{6}{@{}c@{}}{non-linear} \\
\cmidrule{2-7}\cmidrule{8-13}%
& \multicolumn{2}{@{}c@{}}{$\eta_\mu$} & \multicolumn{2}{@{}c@{}}{$\eta_\alpha$} & \multicolumn{2}{@{}c@{}}{$\eta_\pi$} & \multicolumn{2}{@{}c@{}}{$\eta_\mu$} & \multicolumn{2}{@{}c@{}}{$\eta_\alpha$} & \multicolumn{2}{@{}c@{}}{$\eta_\pi$} \\
\cmidrule{2-3}\cmidrule{4-5}\cmidrule{6-7}\cmidrule{8-9}\cmidrule{10-11}\cmidrule{12-13}%
& fixed & opt & fixed & opt & fixed & opt & fixed & opt & fixed & opt & fixed & opt \\ 
\midrule
$x_1$ & \textbf{100} & \textbf{100} & 5 & 13 & 11 & 6 & \textbf{100} & \textbf{100} & 12 & 15 & 10 & 7 \\ 
  $x_2$ & \textbf{100} & \textbf{100} & \textbf{83} & \textbf{94} & 62 & 45 & \textbf{100} & \textbf{100} & \textbf{88} & \textbf{88} & 58 & 53 \\ 
  $x_3$ & \textbf{100} & \textbf{100} & \textbf{94} & \textbf{98} & \textbf{100} & \textbf{100} & \textbf{100} & \textbf{100} & \textbf{97} & \textbf{98} & \textbf{100} & \textbf{100} \\ 
  $x_4$ & \textbf{100} & \textbf{100} & \textbf{78} & \textbf{89} & \textbf{100} & \textbf{100} & \textbf{100} & \textbf{100} & \textbf{92} & \textbf{96} & \textbf{100} & \textbf{100} \\ 
  $x_5$ & 52 & 19 & \textbf{100} & \textbf{100} & \textbf{100} & \textbf{100} & 51 & 25 & \textbf{100} & \textbf{100} & \textbf{100} & \textbf{100} \\ 
  $x_6$ & 54 & 17 & 4 & 11 & \textbf{100} & \textbf{100} & 46 & 15 & 7 & 13 & \textbf{100} & \textbf{100} \\ 
  $x_7$ & 53 & 19 & 2 & 4 & 11 & 6 & 39 & 15 & 2 & 5 & 6 & 6 \\ 
  $x_8$ & 38 & 9 & 7 & 11 & 6 & 4 & 45 & 20 & 4 & 9 & 4 & 3 \\ 
  $x_9$ & 49 & 10 & 5 & 15 & 6 & 1 & 41 & 19 & 5 & 10 & 3 & 3 \\ 
  $x_{10}$ & 53 & 20 & 2 & 7 & 8 & 4 & 55 & 30 & 5 & 6 & 3 & 3 \\ 
  $z_1$ & \textbf{100} & \textbf{100} & \textbf{85} & \textbf{97} & \textbf{100} & \textbf{100} & \textbf{100} & \textbf{100} & \textbf{77} & \textbf{78} & \textbf{100} & \textbf{100} \\ 
  $z_2$ & 72 & 17 & 1 & 2 & 1 & 0 & 82 & 40 & 8 & 20 & 13 & 8 \\ 
\bottomrule
\end{tabular*}
\end{table}

\section{Modeling the Number of Antenatal Care Visits in Nigeria}
\label{daub:section_applications}

\begin{figure}[!b]\centering
\includegraphics[width=0.66\textwidth]{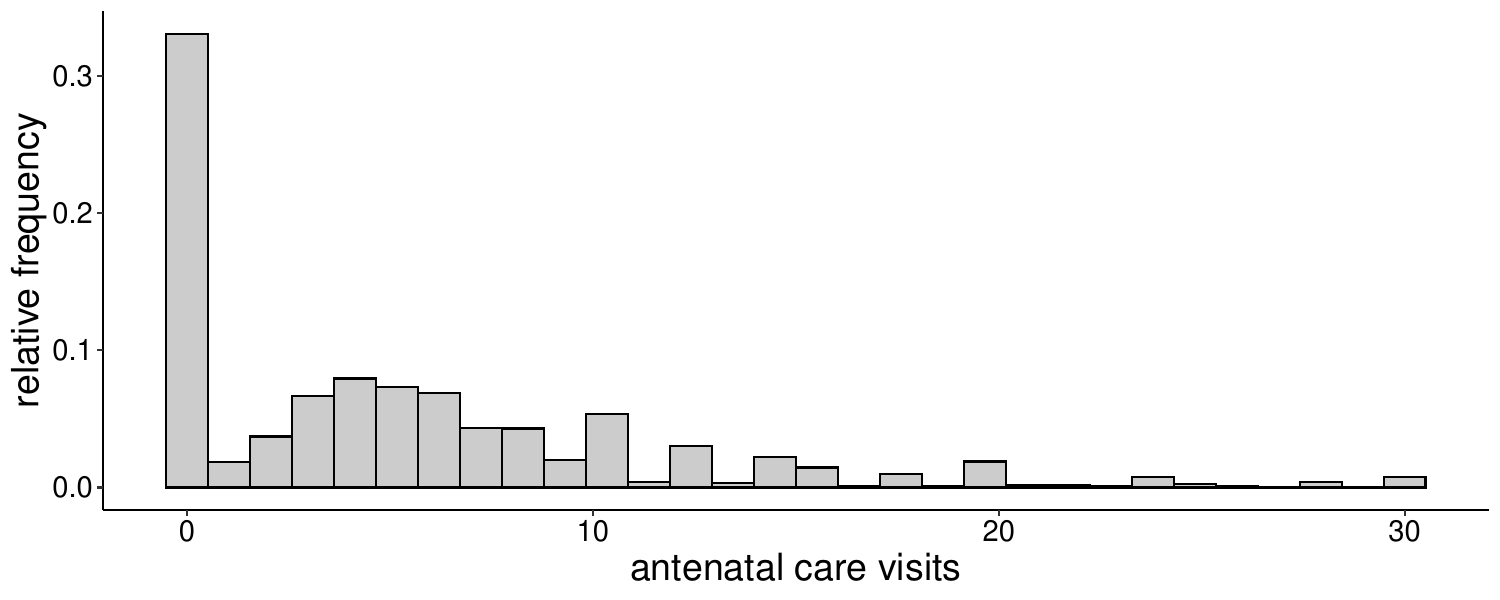} 
\caption{\label{daub:fig_hist_antenatal} Empirical distribution of the number of antenatal care visits}
\end{figure}

In the following, we model the number of antenatal care visits in Nigeria based on maternal socio-economic and demographic characteristics. 
For the analysis we use data from the 2013 Nigeria Demographic and Health Survey \citep{DHSNigeria2013}, 
compiled by \cite{Gayawan2016}. 
The data set contains 18,815 observations of 11 variables, which we split into training and validation set $\left(\frac{2}{3} \text{ vs. } \frac{1}{3}\right)$.
Among the recorded number of antenatal care visits, 34.2\% are zero, while the non-zero observations have a mean of 7.97 and an empirical variance of 34.93 (see Fig.~\ref{daub:fig_hist_antenatal} for the empirical distribution). 
The large proportion of zeros combined with the overdispersion motivates the use of a zero-inflated negative binomial distribution for the response variable.
Aside from the number of antenatal care visits, 
the data set contains different socio-economic, demographic and contextual characteristics of the mother:
\textit{education} (no education, primary school education and higher education), \textit{wealth} (a five-level wealth index), working status (\textit{work}) as well as access to mass media, specifically \textit{newspaper}, \textit{radio} and \textit{television}, represent her economic and social status.
In addition, the mother's age at birth (\textit{age}), the duration of her marriage (\textit{married\_yrs}), her region of residence (\textit{zone}) and whether she lives in an urban or rural area \textit{(urban)} are included. 
Descriptive statistics on the covariates are provided in the Supplement.

\begin{figure}[!b]
\includegraphics[width=\textwidth]
{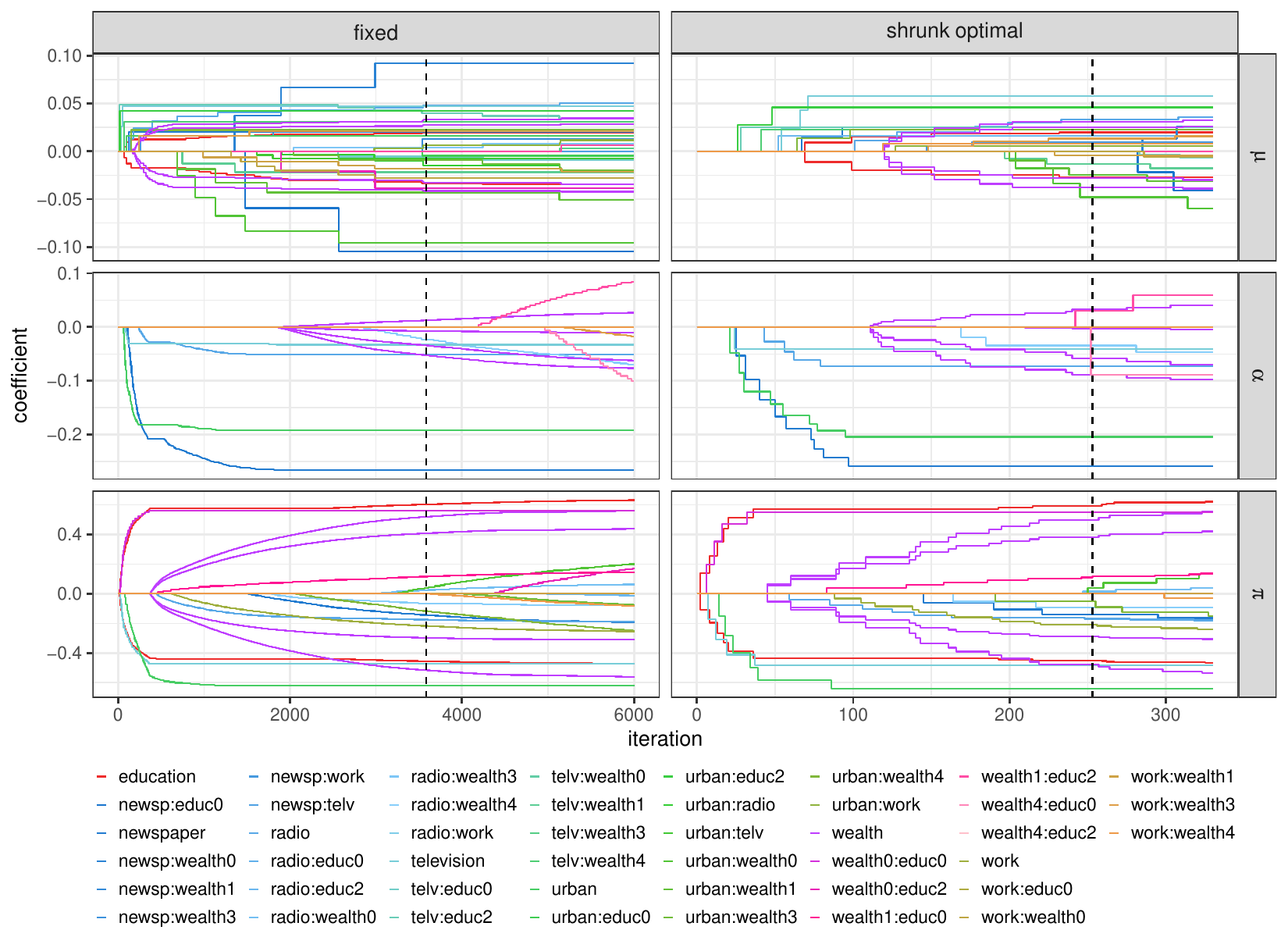}
\caption{\label{daub:fig_pathplot} Coefficient paths for the antenatal care data from Nigeria using different step length approaches (columns). The dashed vertical lines represent the stopping iterations. Note that the x-axis scaling varies across the step lengths approaches}
\end{figure}

Based on the characteristics of the response variable and because we are not only interested in the effects on its conditional mean but also on the other distribution parameters,
a zero-inflated negative binomial model for location, scale and shape is considered to model the number of antenatal care visits.
Aside from the conditional mean, the effects on the zero-inflation parameter are particularly relevant, since having no antenatal care is especially problematic from a medical point of view.
For the estimation of the ZINB-GAMLSS via boosting, the categorical covariates (\textit{education}, \textit{wealth}) are dummy-coded and combined into a single effect,  
the continuous covariates (\textit{age}, \textit{married\_yrs}) are modeled using P-splines decomposed into a linear effect and a smooth deviation, and the discrete spatial effect of \textit{zone} is incorporated via a Markov random field. 
The number of equivalent degrees of freedom is set to two for these base-learners.
The remaining covariates, all of which are binary, 
enter the model as simple linear effects.
In addition to these main effects, we include all pairwise interactions between binary and categorical covariates. 
For the categorical covariates, the interactions are defined at the level of the individual categories, 
reflecting the possibility that interaction effects might only be present for certain category combinations. 
In the following, we compare the results from boosting with fixed and shrunk optimal step lengths, where the fixed step length and the shrinkage factor of the optimal step length are set to 0.1.
The stopping iteration is again determined based on a robust 10-fold cross-validation variant. 
As reference, the boosting results are additionally compared with direct maximum likelihood estimation for a model without interaction terms.
The corresponding estimates are included in the Supplement. 

Fig.~\ref{daub:fig_pathplot} displays the coefficient paths for the two step length approaches. 
When using shrunk optimal step lengths, the paths for $\mu$ grow notably slower relative to the other predictors than in the fixed step length approach, while the paths for $\alpha$ grow faster.
The coefficient estimates of the different submodels, therefore, converge at a similar speed, 
indicating a more balanced update behavior. 
In contrast, with fixed step lengths the convergence rates of the additive predictors differ substantially: the coefficients for $\mu$ converge much faster whereas those for $\alpha$ converge considerably more slowly than the coefficients for $\pi$.
As a result, the coefficient estimates exhibit a different degree of shrinkage across the two step length approaches, where the effects on $\mu$ are often more regularized when using shrunk optimal step lengths.
For example the estimated interaction effects of \textit{newspaper:wealth0}, \textit{newspaper:education0} and \textit{urban:wealth0} are considerably higher for the fixed step length approach (an overview of all estimates is presented in the Supplement). 
Exceptions such as the effect of access to \textit{television} on $\mu$, for which the coefficient of the shrunk optimal approach is larger, 
appear to result from partial compensation of the smaller interaction effects.
In line with the stronger coefficient shrinkage, 
the shrunk optimal step length approach selects ten base-learners less than the fixed step length approach for $\mu$ (see Table~\ref{daub:table_sel_covs_antenatal}). 
The additive predictor associated with $\alpha$, on the other hand, is subject to less regularization when optimal step lengths are used, which becomes apparent in the larger coefficient estimates of \textit{wealth}, \textit{radio} and \textit{television} as well as in the selection of two additional base-learners.
For the zero-inflation parameter $\pi$, both step length approaches yield a similar degree of regularization, with the shrunk optimal step length approach showing a slightly stronger shrinkage.

\begin{table}[!t]\centering
\caption{\label{daub:table_sel_covs_antenatal} Number of selected base-learners per additive predictor for the antenatal care data from Nigeria using different step lengths approaches (rows)}
\medskip
\begin{tabular}{lccc}
\toprule[0.09 em]
 & $\mu$ & $\alpha$ & $\pi$ \\ 
\midrule
fixed &  32 &  8 &  17 \\ 
shrunk optimal &  22 &  10 &  15 \\ 
\bottomrule[0.09 em]
\end{tabular}
\end{table}

\begin{figure}[!b]\centering
\includegraphics[width=\textwidth]
{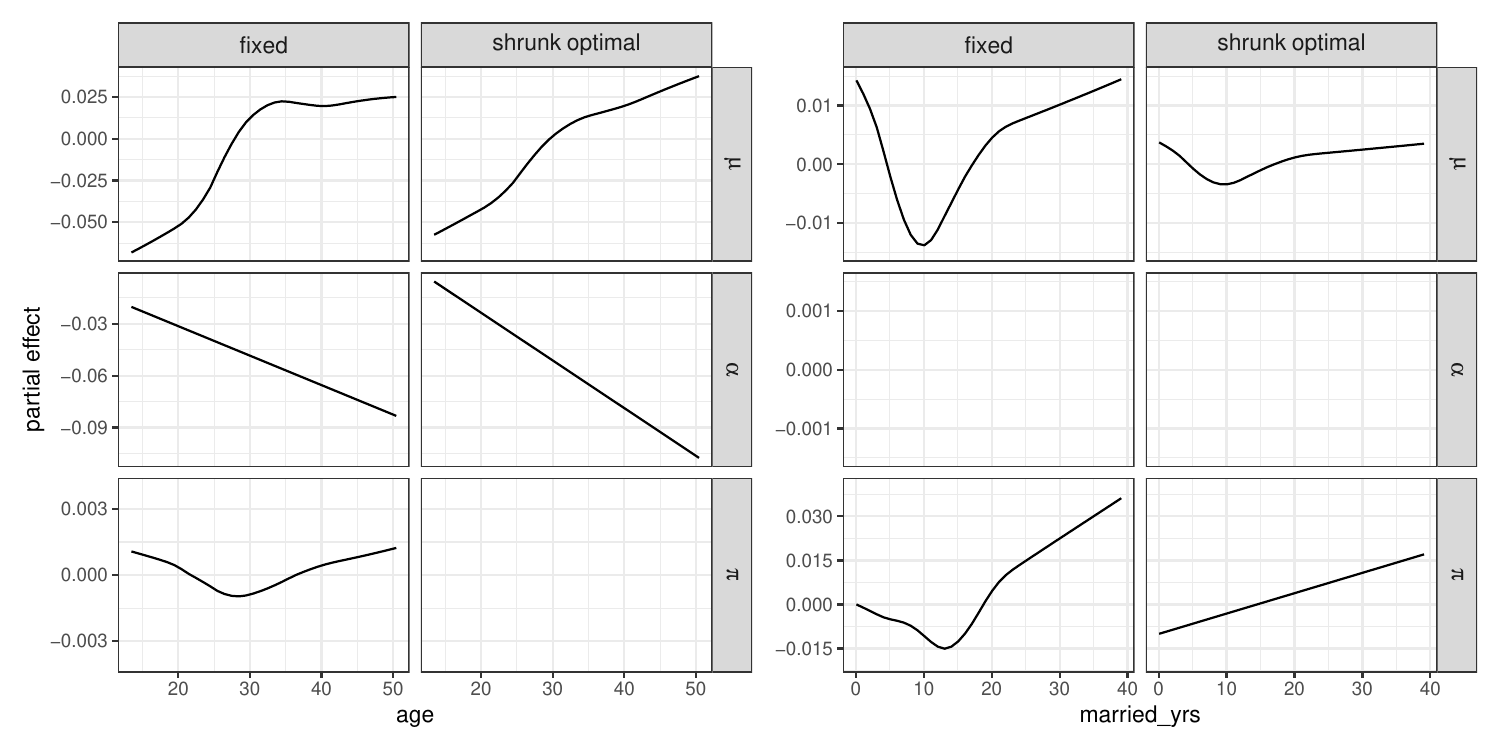} 
\caption{\label{daub:fig_partial_mab_mardur} Partial effects of the mother's age (left) and marriage duration (right) for the antenatal care data from Nigeria using different step length approaches (columns)}
\end{figure}

\begin{figure}[!t]\centering
\includegraphics[width=\textwidth]
{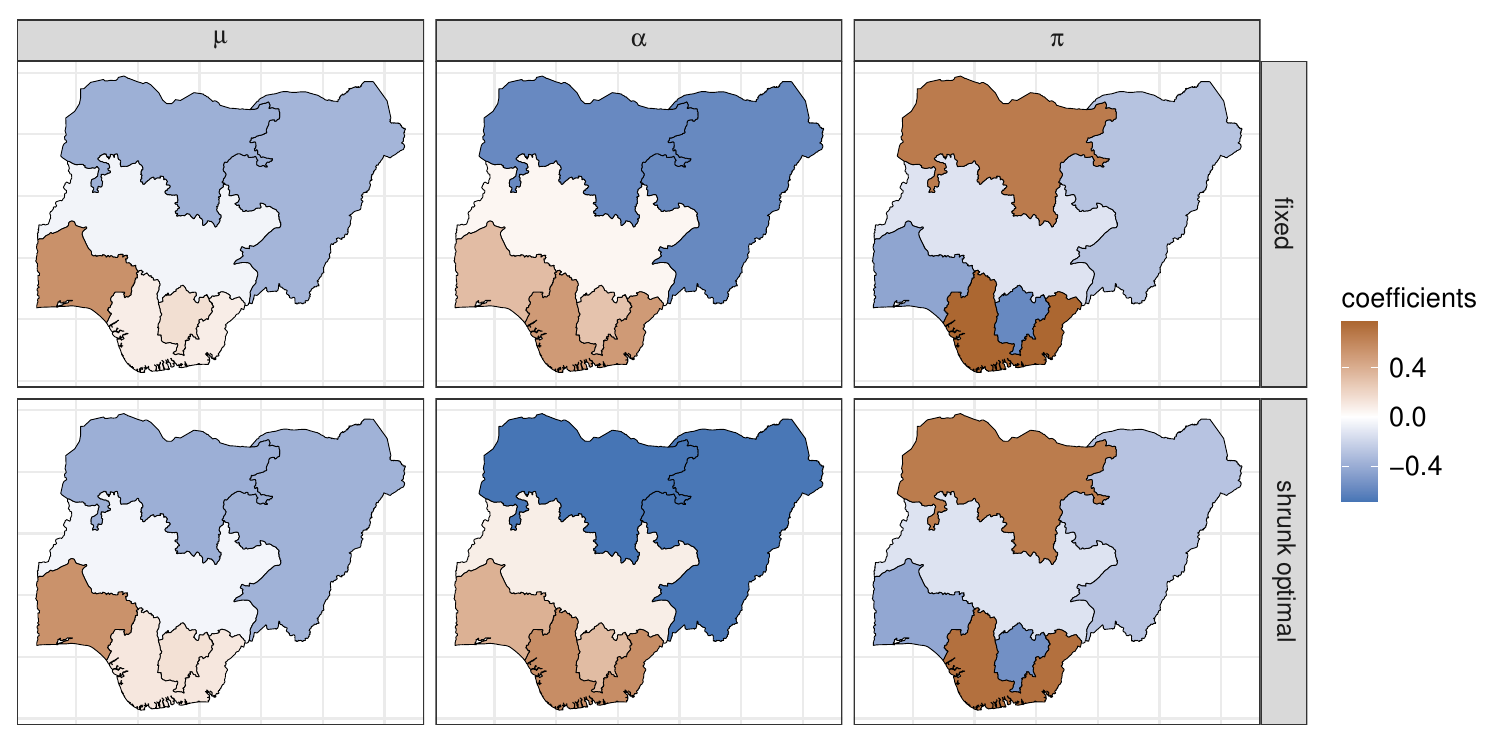} 
\caption{\label{daub:fig_spatial_antenatal} Spatial effects for the antenatal care data from Nigeria using different step length approaches (rows)}
\end{figure}

The differing degrees of regularization are also reflected in the estimates of the non-linear and spatial effects.
When shrunk optimal step lengths are used, the estimated effects of both \textit{age} and \textit{married\_yrs} on $\mu$ are smaller or less flexible. 
For $\alpha$, shrunk optimal step lengths yield a larger effect for \textit{age} while \textit{married\_yrs} is not selected by either approach (see Fig.~\ref{daub:fig_partial_mab_mardur}).
As the effect estimates of both continuous covariates on $\pi$ are comparatively small, the slightly stronger regularization of $\pi$ for the shrunk optimal step length approach results in differences in the covariate and effect selection. 
Specifically, \textit{age} is not selected at all, and the effect of \textit{married\_yrs} is regularized to a linear effect by the shrunk optimal approach, while the fixed step length approach includes both smooth deviations. 
For the spatial effects on $\mu$ and $\pi$, the step length approaches yield a similar degree of regularization, 
whereas the spatial effect on $\alpha$ is again less regularized for shrunk optimal step lengths (see Fig.~\ref{daub:fig_spatial_antenatal}). 
Overall, the shrunk optimal step length approach thus leads to a more balanced model due to the more similar degree of regularization and smaller differences in the number of selected covariates  across predictors. 

\begin{figure}[!t]\centering
\includegraphics[width=\textwidth]
{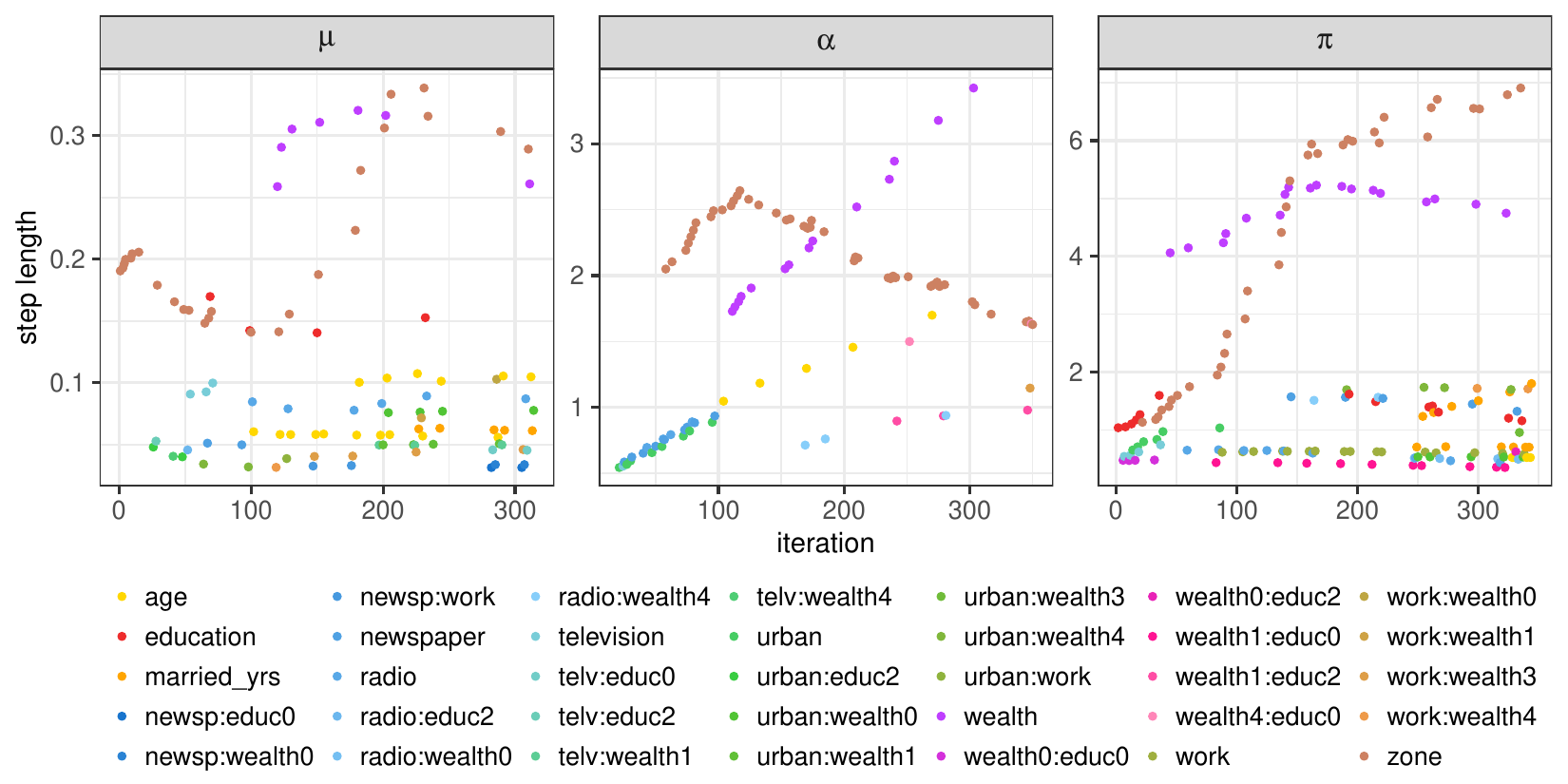} 
\caption{\label{daub:fig_steplengths} Shrunk optimal step lengths for the antenatal care data from Nigeria}
\end{figure}

This improvement in balancedness is driven by substantial differences in the shrunk optimal step lengths across the additive predictors, which are shown in Fig.~\ref{daub:fig_steplengths}. 
For unpenalized base-learners, the optimal step lengths take values around 0.5 for $\mu$, 8 for $\alpha$ and 5 for $\pi$ (for some base-learners larger); 
a constant optimal step length of 1 over all iterations and across all additive predictors would coincide with the fixed step length approach.
Compared to the fixed step length approach, 
the considerably larger step lengths for $\alpha$ and $\pi$ lead to a substantially faster convergence of their additive predictors, whereas the smaller values for $\mu$ slow down its convergence.
This shift in relative convergence behavior results in the more similar convergence speed across additive predictors and thereby contributes to a more balanced model.

Differences in the step length level also arise within the same additive predictor.
Consistent with previous findings, step lengths for penalized base-learners tend to take larger values than step lengths for unpenalized base-learners. 
In particular, step lengths corresponding to the categorical effects of \textit{wealth} and \textit{education} as well as the spatial effect of \textit{zone} exhibit comparatively large values. 
While for example comparing the step lengths of \textit{education} and \textit{wealth} we observe larger optimal step lengths for the stronger penalized base-learners, the relation between the optimal step lengths levels of different base-learners cannot only be reduced to the degree of penalization. 
This is also reflected by the differences in the behavior of optimal step lengths across the additive predictors. 
In addition to an improved relative convergence behavior, shrunk optimal step lengths yield a faster overall convergence, resulting in a considerably earlier stopping iteration ($253$ compared to $3,584$).

In order to compare the predictive performance, we evaluate the CRPS (with the observed value as reference) and the negative log-likelihood on the validation set. 
Both approaches achieve a very similar predictive accuracy with respect to these measures;
however, the shrunk optimal approach attains this performance with a sparser model.
These findings are consistent with the simulation results; 
more detailed results are provided in the Supplement.

Concerning the covariate effects on the distribution of antenatal care visits, we obtain the following results from boosting with shrunk optimal step lengths (see Fig.~\ref{daub:fig_partial_mab_mardur}, Fig.~\ref{daub:fig_spatial_antenatal} and the Supplement for the coefficient table). 
For the mean parameter $\mu$, the largest positive effect can be observed for access to \textit{television} while access to the other mass media (\textit{radio}, \textit{newspaper}) exhibits more moderate positive effects.
Although \textit{urban} itself is not selected, strong effects are apparent for the interactions \textit{urban:education2}, \textit{urban:wealth0} and \textit{urban:wealth1}. 
Moreover, \textit{wealth} and \textit{education} exert substantial effects on $\mu$ and additionally enter the model through interactions; 
aside from the interactions involving \textit{urban} the most notable ones are \textit{television:education2} and \textit{television:wealth4}. 
Also for $\pi$, \textit{education}, \textit{wealth}, \textit{urban} and \textit{television} show large effects on the additive predictor. 
In contrast to $\mu$, these effects primarily arise from the main effects rather than from interactions with other covariates. 
The only interaction with a pronounced effect is \textit{wealth0:education0}, while \textit{wealth1:education0}, \textit{work} as well as access to \textit{radio} and \textit{newspaper} exhibit more moderate effects. 
With respect to the overdispersion parameter $\alpha$, strong negative effects are observed for \textit{newspaper} and \textit{urban}.
Access to the other mass media (\textit{radio}, \textit{telelvision}) as well as \textit{wealth} show moderate effects.

For the mother's \textit{age} at birth we obtain a positive effect on $\mu$, which is slightly more enhanced in the age group 25 to 35, and a negative effect on $\alpha$.
The effect of \textit{married\_yrs} on $\mu$ is considerably smaller compared to the effect of \textit{age} and the direction varies across the range:
for a marriage duration below 10 years the effect on $\mu$ is negative while it is positive for a larger marriage duration.  
With respect to $\pi$, \textit{married\_yrs} has a small positive effect.
Regarding the spatial effects on $\mu$ and $\alpha$, pronounced north-south differences are evident, with the northern zones exhibiting negative and the southern zones positive effects on both parameters. 
For $\pi$, the spatial pattern is more variable: a strong positive effect is evident for the North West and the South South zones, while the remaining zones show more moderate negative effects.

\section{Conclusion and Discussion}
\label{daub:section_conclusion}

Using shrunk optimal step lengths for non-cyclical boosting of GAMLSS has been shown to ensure a balanced overall model by enabling a fairer update selection and leveling the convergence speed of the different additive predictors.
In the present work, we extended this approach to a larger variety of base-learners, considering jointly dummy-coded penalized linear base-learners for modeling categorical effects, P-spline base-learners representing non-linear effects and Markov random field base-learners to account for discrete spatial effects. 
When combined with base-learners that penalize the overall size of the fit, larger optimal step lengths were observed compared to unpenalized base-learners, reflecting the greater potential for reducing the loss function.
This constitutes an additional benefit of shrunk optimal step lengths in this context, as it increases the otherwise slow convergence speed of such effects while maintaining a fair base-learner selection.
Furthermore, we extended the class of GAMLSS boosted with shrunk optimal step lengths to three-parameter models, specifically a zero-inflated negative binomial model for location, scale and shape, motivated by our application. 

Simulation studies and the application further demonstrate that using shrunk optimal step lengths reduces the number of false positives and mitigates differences in regularization across additive predictors compared to the fixed step length approach for effects not represented by simple linear base-learners by a similar magnitude as for simple linear base-learners.
In the case of a zero-inflated negative binomial response variable, relative regularization of the additive predictors and variable selection behavior were consistent with those observed for other response variable distributions. 
As opposed to the Gaussian case, the predictive performance is not improved by using shrunk optimal step lengths here, but a similar performance is achieved with a sparser model. 

The analysis of determinants of antenatal care utilization indicates that wealth, education, access to television, type of residential area and region of residence exert the strongest effects on the conditional mean and the zero-inflation probability.
For the conditional mean, interactions between these covariates have a large impact, whereas the zero-inflation parameter is primarily driven by the main effects. 
The overdispersion parameter is most strongly influenced by access to newspaper, type of residential area and the region of residence.

In summary, shrunk optimal step lengths can readily be applied to base-learners beyond simple linear ones as well as to a zero-inflated negative binomial response variable distribution, where they improve variable selection, lead to a more balanced regularization across additive predictors and enhance estimation efficiency.
In none of the considered settings did shrunk optimal step lengths lead to overall inferior results compared to the fixed step length approach. 

Some limitations inherent to boosting methods remain, which are not related to the step length approach.
While the use of shrunk optimal step lengths mitigates the lower relative importance of additive predictors associated with small negative gradients, such predictors can nevertheless be challenging to estimate.
This is, for instance, evident for the overdispersion parameter in the ZINB-GAMLSS.
Further research is therefore warranted to be able to better address this issue.
In addition, a comparatively large number of false positives is selected across the additive predictors, which reflects a well-known characteristic of boosting-based approaches. 
Alternative stopping criteria, such as probing \citep{Thomas2017}, or a deselection procedure that removes covariate effects of minor importance a posteriori \citep{Stroemer2022} could be employed in future work to address this issue.
Another inherent limitation of boosting methods is the lack of standard errors for the estimated effects and related inferential statistics.
To obtain these quantities, resampling-based approaches such as bootstrapping could for example be applied. 

Our implementation of the boosting approach with shrunk optimal step lengths, building on the R package \textit{gamboostLSS}, is publicly available on GitHub ({\urlstyle{same}\url{https://github.com/AlexandraDaub/BoostZINB}}). 
As a next step, we plan to integrate the option of using shrunk optimal step lengths directly into the \textit{gamboostLSS} package to facilitate application in practice. 
Future work will further explore the extensions of this approach to additional base-learner types as well as to a wider range of response variable distributions.
Moreover, an interesting direction for future research is the application of boosting with shrunk optimal step lengths to more complex model classes within the GAMLSS framework, such as distributional copula regression \citep{Hans2022} or multivariate distributional regression \citep{Stroemer2023}.
Lastly, the potential benefits of using shrunk optimal step lengths in model classes with a single predictor remain to be explored, particularly given that our findings suggest that optimal step lengths may also be advantageous for boosting generalized additive models in settings with certain types of penalized base-learners.

\subsection*{Acknowledgements} 
The authors would like to thank Ezra Gayawan for providing the data and for valuable input regarding the consistency and interpretation of the results.

\subsection*{Funding} 
This work was supported by the DFG (Deutsche Forschungsgemeinschaft; Projekt BE 7939/2-2).

\bibliographystyle{apalike} 
\bibliography{literature_paper2}

\newpage

\appendix
\noindent {\Huge \textbf{Supplement} }
\setcounter{page}{1}

\section{Derivation of the first-order condition for optimal step lengths in ZINB-GAMLSS}

For a ZINB-GAMLSS, we have a log-likelihood of 
\begin{align*}
\sum_{i=1}^n &\ell\left(\mu_i, \alpha_i, \pi_i; y_i \right) \\
&= \sum_{i: y_i=0}  \ln\left(\pi_i + (1-\pi_i)(1+\alpha_i \mu_i)^{-\frac{1}{\alpha_i}}\right) \\
&\hspace*{1em}+ \sum_{i: y_i>0} \ln(1-\pi_i) + y_i \ln \left( \frac{\alpha_i \mu_i}{1+\alpha_i \mu_i}\right) - \frac{1}{\alpha_i} \ln(1+\alpha_i \mu_i) + \ln\Gamma\left(y_i+\frac{1}{\alpha_i}\right) - \ln\Gamma(y_i+1) - \ln\Gamma\left(\frac{1}{\alpha_i}\right),
\end{align*}
where $\Gamma$ denotes the gamma function. 

\subsection{Derivation of the first-order condition of $\boldsymbol{\nu_\mu^*}$}

In order to determine the optimal step length $\nu_\mu^{*}$ in iteration $m$, the following optimization problem 
has to be solved
\begin{align}
\nu^{*[m]}_\mu = \underset{\nu_\mu}{\text{arg min}} - \sum_{i=1}^n \ell\left(\exp\left(\hat{\eta}_{\mu,i}^{[m-1]} + \nu_\mu \hat{h}_{\mu,i}^{*[m]}\right), \hat{\alpha}^{[m-1]}_i, \hat{\pi}^{[m-1]}_i; y_i \right),
\label{daub:formula_suppl_opt_problem_nu_mu}
\end{align}
where for notational convenience we define here and in the following $\hat{\eta}_{\mu,i}^{[m]} = \hat{\eta}_\mu^{[m]}(\boldsymbol{x}_i)$, $\hat{h}_{\mu,i}^{*[m]} = \hat{h}_\mu^{*[m]}(\boldsymbol{x}_i)$, $\hat{\mu}_i^{[m]} = \hat{\mu}^{[m]}(\boldsymbol{x}_i)$, $\hat{\alpha}_i^{[m]} = \hat{\alpha}^{[m]}(\boldsymbol{x}_i)$ and $\hat{\pi}_i^{[m]} = \hat{\pi}^{[m]}(\boldsymbol{x}_i)$. 
For the first order condition of (\ref{daub:formula_suppl_opt_problem_nu_mu}) holds
\begin{align*}
0 &= - \sum_{i=1}^n \frac{\partial}{\partial \nu_\mu} \ell\left(\exp\left(\hat{\eta}_{\mu,i}^{[m-1]} + \nu_\mu \hat{h}_{\mu,i}^{*[m]}\right), \hat{\alpha}^{[m-1]}_i, \hat{\pi}^{[m-1]}_i; y_i \right) \\
&= - \sum_{i=1}^n \left( \frac{\partial \ell}{\partial \mu^{[m]}_i}\right) \left(\frac{\partial \mu^{[m]}_i}{\partial \eta_{\mu,i}^{[m]}}\right) \left(\frac{\partial \eta_{\mu,i}^{[m]}}{\partial \nu_\mu}\right).
\end{align*}
With $\mu^{[m]}_i = \exp\left(\eta^{[m]}_{\mu,i}\right)$ and $\eta^{[m]}_{\mu,i} = \eta_{\mu,i}^{[m-1]} + \nu_\mu h_{\mu,i}^{*[m]}$, we obtain for the partial derivatives:
\begin{align*}
\frac{\partial \ell}{\partial \mu^{[m]}_i}
&= \begin{cases} 
\displaystyle
\frac{-\left(1-\pi^{[m-1]}_i\right) \left(1 + \alpha^{[m-1]}_i \mu^{[m]}_i\right)^{-\frac{1}{\alpha^{[m-1]}_i}-1}}
{\pi^{[m-1]}_i + \left(1-\pi^{[m-1]}_i\right) \left(1 + \alpha^{[m-1]}_i \mu^{[m]}_i \right)^{-\frac{1}{\alpha^{[m-1]}_i}}} \quad &\text{if } y_i = 0\\
\displaystyle
\frac{y_i}{\mu^{[m]}_i \left(1 + \alpha^{[m-1]}_i \mu^{[m]}_i\right)} - \frac{1}{1 + \alpha^{[m-1]}_i \mu^{[m]}_i} \quad &\text{if } y_i > 0
\end{cases} \\
&= 
\begin{cases} 
\displaystyle
\frac{-1}
{\frac{\pi^{[m-1]}_i}{1-\pi^{[m-1]}_i} \left(1 + \alpha^{[m-1]}_i \mu^{[m]}_i\right)^{\frac{1}{\alpha^{[m-1]}_i}+1} + \left(1 + \alpha^{[m-1]}_i \mu^{[m]}_i \right)} \quad &\text{if } y_i = 0 \\
\displaystyle
\frac{y_i - \mu^{[m]}_i}{\mu^{[m]}_i \left(1 + \alpha^{[m-1]}_i \mu^{[m]}_i\right)} \quad &\text{if } y_i > 0
\end{cases} \\
\frac{\partial \mu^{[m]}_i}{\partial \eta^{[m]}_{\mu,i}} &= \exp\left(\eta^{[m]}_{\mu,i}\right) = \mu^{[m]}_i \\
\frac{\partial \eta^{[m]}_{\mu,i}}{\partial \nu_\mu} &= h_{\mu,i}^{*[m]}
\end{align*}
The first order condition of (\ref{daub:formula_suppl_opt_problem_nu_mu}) thus is
\begin{align*}
0 &= - \frac{\partial}{\partial \nu_\mu} \sum_{i=1}^n \ell\left(\exp\left(\hat{\eta}_{\mu,i}^{[m-1]} + \nu_\mu \hat{h}_{\mu,i}^{*[m]}\right), \hat{\alpha}^{[m-1]}_i, \hat{\pi}^{[m-1]}_i; y_i \right) \Bigg\vert_{\nu_\mu=\nu_\mu^{*[m]}} \\
&= - \sum_{i:y_i=0} \frac{-\mu^{[m]}_i h_{\mu,i}^{*[m]}}
{\frac{\pi^{[m-1]}_i}{1-\pi^{[m-1]}_i} \left(1 + \alpha^{[m-1]}_i \mu^{[m]}_i\right)^{\frac{1}{\alpha^{[m-1]}_i}+1} + \left(1 + \alpha^{[m-1]}_i \mu^{[m]}_i \right)} \\
&\hspace{2em} - \sum_{i:y_i>0} \frac{\mu^{[m]}_i h_{\mu,i}^{*[m]} \left(y_i - \mu^{[m]}_i \right)}
{\mu^{[m]}_i \left(1 + \alpha^{[m-1]}_i \mu^{[m]}_i\right)} \\
&= \sum_{i:y_i=0} \frac{\exp\left(\eta_{\mu,i}^{[m-1]} + \nu_\mu^{*[m]} h_{\mu,i}^{*[m]}\right) h_{\mu,i}^{*[m]}}
{\frac{\pi^{[m-1]}_i}{1-\pi^{[m-1]}_i} \left(1 + \alpha^{[m-1]}_i \exp\left(\eta_{\mu,i}^{[m-1]} + \nu_\mu^{*[m]} h_{\mu,i}^{*[m]}\right)\right)^{\frac{1}{\alpha^{[m-1]}_i}+1} + \left(1 + \alpha^{[m-1]}_i \exp\left(\eta_{\mu,i}^{[m-1]} + \nu_\mu^{*[m]} h_{\mu,i}^{*[m]}\right) \right)} \\
&\hspace{2em} - \sum_{i:y_i>0} \frac{h_{\mu,i}^{*[m]} \left(y_i - \exp\left(\eta_{\mu,i}^{[m-1]} + \nu_\mu^{*[m]} h_{\mu,i}^{*[m]}\right)\right)}{1 + \alpha_i^{[m-1]} \exp\left(\eta_{\mu,i}^{[m-1]} + \nu_\mu^{*[m]} h_{\mu,i}^{*[m]}\right)} .
\end{align*}


\subsection{Derivation of the first-order condition of $\boldsymbol{\nu_\alpha^*}$}
In order to determine the optimal step length $\nu_\alpha^{*}$ in iteration $m$, the following optimization problem has to be solved
\begin{align}
\nu^{*[m]}_\alpha = \underset{\nu_\alpha}{\text{arg min}} - \sum_{i=1}^n \ell\left(\hat{\mu}^{[m-1]}_i, \exp\left(\hat{\eta}_{\alpha,i}^{[m-1]} + \nu_\alpha \hat{h}_{\alpha,i}^{*[m]}\right), \hat{\pi}^{[m-1]}_i; y_i \right).
\label{daub:formula_suppl_opt_problem_nu_alpha}
\end{align}
For the first order condition of (\ref{daub:formula_suppl_opt_problem_nu_alpha}) holds
\begin{align*}
0 &= - \sum_{i=1}^n \frac{\partial \ell}{\partial \nu_\alpha} \left(\hat{\mu}^{[m-1]}_i, \exp\left(\hat{\eta}_{\alpha,i}^{[m-1]} + \nu_\alpha \hat{h}_{\alpha,i}^{*[m]}\right), \hat{\pi}^{[m-1]}_i; y_i \right) \\
&= - \sum_{i=1}^n \left( \frac{\partial \ell}{\partial \alpha^{[m]}_i}\right) \left(\frac{\partial \alpha^{[m]}_i}{\partial \eta_{\alpha,i}^{[m]}}\right) \left(\frac{\partial \eta_{\alpha,i}^{[m]}}{\partial \nu_\alpha}\right)
\end{align*}
With $\alpha^{[m]}_i = \exp\left(\eta^{[m]}_{\alpha,i}\right)$ and $\eta^{[m]}_{\alpha,i} = \eta_{\alpha,i}^{[m-1]} + \nu_\alpha h_{\alpha,i}^{*[m]}$, we obtain for the partial derivatives
\begin{align*}
\frac{\partial \ell}{\partial \alpha^{[m]}_i}
&= \begin{cases} 
\displaystyle
\frac{\left(1-\pi^{[m-1]}_i\right) \left(1+\alpha^{[m]}_i \mu^{[m-1]}_i \right)^{-\frac{1}{\alpha^{[m]}_i}} 
\left[\left(-\frac{1}{\alpha^{[m]}_i}\right) \frac{\mu^{[m-1]}_i}{1+\alpha^{[m]}_i \mu^{[m-1]}_i} + \frac{1}{{\alpha^{[m]}_i}^2} \ln\left(1+\alpha^{[m]}_i \mu^{[m-1]}_i\right)\right]}
{\pi^{[m-1]}_i + \left(1-\pi^{[m-1]}_i\right) \left(1+\alpha^{[m]}_i \mu^{[m-1]}_i \right)^{-\frac{1}{\alpha^{[m]}_i}}} \quad &\text{if } y_i = 0\\
\displaystyle
y_i \left(\frac{\mu^{[m-1]}_i}{\alpha^{[m]}_i \mu^{[m-1]}_i} - \frac{\mu^{[m-1]}_i}{1+\alpha^{[m]}_i \mu^{[m-1]}_i}\right) 
- \frac{\mu^{[m-1]}_i}{\alpha^{[m]}_i \left(1+\alpha^{[m]}_i \mu^{[m-1]}_i\right)} + \frac{1}{{\alpha^{[m]}_i}^2} \ln\left(1+\alpha^{[m]}_i \mu^{[m-1]}_i \right)  \\
\displaystyle
\quad+ \psi\left(y_i + \frac{1}{\alpha^{[m]}_i} \right) \left(- \frac{1}{{\alpha^{[m]}_i}^2}\right)
- \psi\left(\frac{1}{\alpha^{[m]}_i} \right) \left(- \frac{1}{{\alpha^{[m]}_i}^2}\right)
\quad &\text{if } y_i > 0
\end{cases} \\
&= 
\begin{cases} 
\displaystyle
\frac{\frac{1}{\alpha^{[m]}_i} \left(\frac{1}{\alpha^{[m]}_i} \ln\left(1+\alpha^{[m]}_i \mu^{[m-1]}_i\right) - \frac{\mu^{[m-1]}_i}{1+\alpha^{[m]}_i \mu^{[m-1]}_i}\right)}
{\frac{\pi^{[m-1]}_i}{1-\pi^{[m-1]}_i} \left(1+\alpha^{[m]}_i \mu^{[m-1]}_i \right)^{\frac{1}{\alpha^{[m]}_i}} + 1} \quad &\text{if } y_i = 0 \\
\displaystyle
\frac{y_i - \mu^{[m-1]}_i}{\alpha^{[m]}_i \left(1+\alpha^{[m]}_i \mu^{[m-1]}_i\right)} 
+ \frac{1}{{\alpha^{[m]}_i}^2} \left[ \ln\left(1+\alpha^{[m]}_i \mu^{[m-1]}_i \right) - \psi\left(y_i + \frac{1}{\alpha^{[m]}_i} \right) + \psi\left(\frac{1}{\alpha^{[m]}_i} \right)\right] \quad &\text{if } y_i > 0
\end{cases} \\
\frac{\partial \alpha^{[m]}_i}{\partial \eta^{[m]}_{\alpha,i}} &= \exp\left(\eta^{[m]}_{\alpha,i}\right) = \alpha^{[m]}_i \\
\frac{\partial \eta^{[m]}_{\alpha,i}}{\partial \nu_\alpha} &= h_{\alpha,i}^{*[m]},
\end{align*}
where $\psi$ denotes the digamma function. \\

\noindent The first order condition of (\ref{daub:formula_suppl_opt_problem_nu_alpha}) thus is
\begin{align*}
0 &= - \frac{\partial}{\partial \nu_\alpha} \sum_{i=1}^n \ell\left(\hat{\mu}^{[m-1]}_i, \exp\left(\hat{\eta}_{\alpha,i}^{[m-1]} + \nu_\alpha \hat{h}_{\alpha,i}^{*[m]}\right), \hat{\pi}^{[m-1]}_i; y_i \right) \Bigg\vert_{\nu_\alpha=\nu_\alpha^{*[m]}} \\
&= - \sum_{i:y_i=0} \frac{\frac{1}{\alpha^{[m]}_i} \left(\frac{1}{\alpha^{[m]}_i} \ln\left(1+\alpha^{[m]}_i \mu^{[m-1]}_i\right) - \frac{\mu^{[m-1]}_i}{1+\alpha^{[m]}_i \mu^{[m-1]}_i}\right)}
{\frac{\pi^{[m-1]}_i}{1-\pi^{[m-1]}_i} \left(1+\alpha^{[m]}_i \mu^{[m-1]}_i \right)^{\frac{1}{\alpha^{[m]}_i}} + 1} \alpha^{[m]}_i h_{\alpha,i}^{*[m]}  \\
&\hspace{2em} - \sum_{i:y_i>0} \left( \frac{y_i - \mu^{[m-1]}_i}{\alpha^{[m]}_i \left(1+\alpha^{[m]}_i \mu^{[m-1]}_i\right)} 
+ \frac{1}{{\alpha^{[m]}_i}^2} \left[ \ln\left(1+\alpha^{[m]}_i \mu^{[m-1]}_i \right) - \psi\left(y_i + \frac{1}{\alpha^{[m]}_i} \right) + \psi\left(\frac{1}{\alpha^{[m]}_i} \right)\right] \right) \alpha^{[m]}_i h_{\alpha,i}^{*[m]}  \\
&= - \sum_{i:y_i=0} \frac{h_{\alpha,i}^{*[m]} \left(\frac{\ln\left(1+\alpha^{[m]}_i \mu^{[m-1]}_i\right)}{\alpha^{[m]}_i} - \frac{\mu^{[m-1]}_i}{1+\alpha^{[m]}_i \mu^{[m-1]}_i}\right)}
{\frac{\pi^{[m-1]}_i}{1-\pi^{[m-1]}_i} \left(1+\alpha^{[m]}_i \mu^{[m-1]}_i \right)^{\frac{1}{\alpha^{[m]}_i}} + 1} \\
&\hspace{2em} - \sum_{i:y_i>0} \frac{h_{\alpha,i}^{*[m]} \left(y_i - \mu^{[m-1]}_i\right)}{1+\alpha^{[m]}_i \mu^{[m-1]}_i} 
+ \frac{h_{\alpha,i}^{*[m]} \ln\left(1+\alpha^{[m]}_i \mu^{[m-1]}_i \right)}{\alpha^{[m]}_i} 
+ \frac{h_{\alpha,i}^{*[m]}}{\alpha^{[m]}_i} \left[ - \psi\left(y_i + \frac{1}{\alpha^{[m]}_i} \right) + \psi\left(\frac{1}{\alpha^{[m]}_i} \right)\right] \\
&= - \sum_{i:y_i=0} \frac{h_{\alpha,i}^{*[m]} \left(\frac{\ln\left(1+\mu^{[m-1]}_i \exp\left(\eta_{\alpha,i}^{[m-1]} + \nu_\alpha^{*[m]} h_{\alpha,i}^{*[m]}\right)\right)}{\exp\left(\eta_{\alpha,i}^{[m-1]} + \nu_\alpha^{*[m]} h_{\alpha,i}^{*[m]}\right)} - \frac{\mu^{[m-1]}_i}{1+\mu^{[m-1]}_i \exp\left(\eta_{\alpha,i}^{[m-1]} + \nu_\alpha^{*[m]} h_{\alpha,i}^{*[m]}\right)}\right)}
{\frac{\pi^{[m-1]}_i}{1-\pi^{[m-1]}_i} \left(1+\mu^{[m-1]}_i \exp\left(\eta_{\alpha,i}^{[m-1]} + \nu_\alpha^{*[m]} h_{\alpha,i}^{*[m]}\right) \right)^{\frac{1}{\exp\left(\eta_{\alpha,i}^{[m-1]} + \nu_\alpha^{*[m]} h_{\alpha,i}^{*[m]}\right)}} + 1} \\
&\hspace{2em} - \sum_{i:y_i>0} \frac{h_{\alpha,i}^{*[m]} \left(y_i - \mu^{[m-1]}_i\right)}{1+\mu^{[m-1]}_i\exp\left(\eta_{\alpha,i}^{[m-1]} + \nu_\alpha^{*[m]} h_{\alpha,i}^{*[m]}\right)} 
+ \frac{h_{\alpha,i}^{*[m]} \ln\left(1+\mu^{[m-1]}_i \exp\left(\eta_{\alpha,i}^{[m-1]} + \nu_\alpha^{*[m]} h_{\alpha,i}^{*[m]}\right) \right)}{\exp\left(\eta_{\alpha,i}^{[m-1]} + \nu_\alpha^{*[m]} h_{\alpha,i}^{*[m]}\right)}  \\
&\hspace*{4em}+ \frac{h_{\alpha,i}^{*[m]}}{\exp\left(\eta_{\alpha,i}^{[m-1]} + \nu_\alpha^{*[m]} h_{\alpha,i}^{*[m]}\right)} \left[- \psi\left(y_i + \frac{1}{\exp\left(\eta_{\alpha,i}^{[m-1]} + \nu_\alpha^{*[m]} h_{\alpha,i}^{*[m]}\right)} \right) + \psi\left(\frac{1}{\exp\left(\eta_{\alpha,i}^{[m-1]} + \nu_\alpha^{*[m]} h_{\alpha,i}^{*[m]}\right)} \right)\right].
\end{align*}

\subsection{Derivation of the first-order condition of $\boldsymbol{\nu_\pi^*}$}

In order to determine the optimal step length $\nu_\pi^{*}$ in iteration $m$, the following optimization problem 
has to be solved
\begin{align}
\nu^{*[m]}_\pi = \underset{\nu_\pi}{\text{arg min}} - \sum_{i=1}^n \ell\left(\hat{\mu}^{[m-1]}_i, \hat{\alpha}^{[m-1]}_i, \text{logit}^{-1} \left(\hat{\eta}_{\pi,i}^{[m-1]} + \nu_\pi \hat{h}_{\pi,i}^{*[m]}\right); y_i \right).
\label{daub:formula_suppl_opt_problem_nu_pi}
\end{align}
For the first order condition of (\ref{daub:formula_suppl_opt_problem_nu_pi}) holds
\begin{align*}
0 &= - \sum_{i=1}^n \frac{\partial \ell}{\partial \nu_\pi} \left(\hat{\mu}^{[m-1]}_i, \hat{\alpha}^{[m-1]}_i, \text{logit}^{-1} \left(\hat{\eta}_{\pi,i}^{[m-1]} + \nu_\pi \hat{h}_{\pi,i}^{*[m]}\right); y_i \right) \\
&= - \sum_{i=1}^n \left( \frac{\partial \ell}{\partial \pi^{[m]}_i}\right) \left(\frac{\partial \pi^{[m]}_i}{\partial \eta_{\pi,i}^{[m]}}\right) \left(\frac{\partial \eta_{\pi,i}^{[m]}}{\partial \nu_\pi}\right)
\end{align*}
With $\pi^{[m]}_i = \frac{\exp\left(\eta^{[m]}_{\pi,i}\right)}{1+\exp\left(\eta^{[m]}_{\pi,i}\right)})$ and $\eta^{[m]}_{\pi,i} = \eta_{\pi,i}^{[m-1]} + \nu_\pi h_{\pi,i}^{*[m]}$, we obtain for the partial derivatives:
\newpage
\vspace*{-3.5em}
\begin{align*}
\frac{\partial \ell}{\partial \pi^{[m]}_i}
&= \begin{cases} 
\displaystyle
\frac{1 - \left(1+\alpha^{[m-1]}_i \mu^{[m-1]}_i \right)^{-\frac{1}{\alpha^{[m-1]}_i}}}
{\pi^{[m]}_i + \left(1-\pi^{[m]}_i\right) \left(1+\alpha^{[m-1]}_i \mu^{[m-1]}_i \right)^{-\frac{1}{\alpha^{[m-1]}_i}}} \quad &\text{if } y_i = 0\\
\displaystyle
\frac{-1}{1-\pi^{[m]}_i} \quad &\text{if } y_i > 0
\end{cases} \\
&= 
\begin{cases} 
\displaystyle
\frac{1}{\frac{\left(1+\alpha^{[m-1]}_i \mu^{[m-1]}_i \right)^{-\frac{1}{\alpha^{[m-1]}_i}}}
{1-\left(1+\alpha^{[m-1]}_i \mu^{[m-1]}_i \right)^{-\frac{1}{\alpha^{[m-1]}_i}}} + \pi^{[m]}_i} \quad &\text{if } y_i = 0 \\
\displaystyle
\frac{-1}{1-\pi^{[m]}_i}  \quad &\text{if } y_i > 0
\end{cases} \\
\frac{\partial \pi^{[m]}_i}{\partial \eta^{[m]}_{\pi,i}} &= \frac{\exp\left(\eta^{[m]}_{\pi,i}\right)}{1+\exp\left(\eta^{[m]}_{\pi,i}\right)} - \left(\frac{\exp\left(\eta^{[m]}_{\pi,i}\right)}{1+\exp\left(\eta^{[m]}_{\pi,i}\right)}\right)^2 = \pi^{[m]}_i \left( 1- \pi^{[m]}_i\right) \\
\frac{\partial \eta^{[m]}_{\pi,i}}{\partial \nu_\pi} &= h_{\pi,i}^{*[m]}
\end{align*}
The first order condition of (\ref{daub:formula_suppl_opt_problem_nu_pi}) thus is
\begin{align*}
0 &= - \frac{\partial}{\partial \nu_\pi} \sum_{i=1}^n \ell\left(\hat{\mu}^{[m-1]}_i, \hat{\alpha}^{[m-1]}_i, \text{logit}^{-1} \left(\hat{\eta}_{\pi,i}^{[m-1]} + \nu_\pi \hat{h}_{\pi,i}^{*[m]}\right); y_i \right) \Bigg\vert_{\nu_\pi=\nu_\pi^{*[m]}} \\
&= - \sum_{i:y_i=0} \left(\frac{1}{\frac{\left(1+\alpha^{[m-1]}_i \mu^{[m-1]}_i \right)^{-\frac{1}{\alpha^{[m-1]}_i}}}
{1-\left(1+\alpha^{[m-1]}_i \mu^{[m-1]}_i \right)^{-\frac{1}{\alpha^{[m-1]}_i}}} + \pi^{[m]}_i} \right) \pi^{[m]}_i \left( 1- \pi^{[m]}_i\right) h_{\pi,i}^{*[m]}  \\
&\hspace{2em} - \sum_{i:y_i>0} \frac{-1}{1-\pi^{[m]}_i} \pi^{[m]}_i \left( 1- \pi^{[m]}_i\right) h_{\pi,i}^{*[m]}   \\
&= \sum_{i:y_i=0} \frac{\left( 1- \pi^{[m]}_i\right) h_{\pi,i}^{*[m]} }{\frac{\left(1+\alpha^{[m-1]}_i \mu^{[m-1]}_i \right)^{-\frac{1}{\alpha^{[m-1]}_i}}}
{\pi^{[m]}_i\left(1-\left(1+\alpha^{[m-1]}_i \mu^{[m-1]}_i \right)^{-\frac{1}{\alpha^{[m-1]}_i}}\right)} + 1}  \\
&\hspace{2em} - \sum_{i:y_i>0} \pi^{[m]}_i h_{\pi,i}^{*[m]} \\
&=\sum_{i:y_i=0} \frac{\left( \frac{1}{1+\exp\left(\eta^{[m]}_{\pi,i}\right)}\right) h_{\pi,i}^{*[m]} }{\left(\exp\left(-\eta^{[m]}_{\pi,i}\right) + 1\right) \frac{\left(1+\alpha^{[m-1]}_i \mu^{[m-1]}_i \right)^{-\frac{1}{\alpha^{[m-1]}_i}}}
{\left(1-\left(1+\alpha^{[m-1]}_i \mu^{[m-1]}_i \right)^{-\frac{1}{\alpha^{[m-1]}_i}}\right)} + 1}  \\
&\hspace{2em} - \sum_{i:y_i>0} \left(1 - \frac{1}{1+\exp\left(\eta^{[m]}_{\pi,i}\right)}\right) h_{\pi,i}^{*[m]} \\
&= \sum_{i:y_i=0} \frac{\left( \frac{1}{1+\exp\left(\eta_{\pi,i}^{[m-1]} + \nu_\pi^{*[m]} h_{\pi,i}^{*[m]}\right)}\right) h_{\pi,i}^{*[m]} }{\left(\exp\left(-\eta_{\pi,i}^{[m-1]} - \nu_\pi^{*[m]} h_{\pi,i}^{*[m]}\right) + 1\right) \frac{\left(1+\alpha^{[m-1]}_i \mu^{[m-1]}_i \right)^{-\frac{1}{\alpha^{[m-1]}_i}}}
{\left(1-\left(1+\alpha^{[m-1]}_i \mu^{[m-1]}_i \right)^{-\frac{1}{\alpha^{[m-1]}_i}}\right)} + 1}  \\
&\hspace{2em} - \sum_{i:y_i>0} \left(1 - \frac{1}{1+\exp\left(\eta_{\pi,i}^{[m-1]} + \nu_\pi^{*[m]} h_{\pi,i}^{*[m]}\right)}\right) h_{\pi,i}^{*[m]}.
\end{align*}

\newpage
\section{Simulation}

\subsection{Gaussian setting}
\label{daub:appdx_gaussian_setting}
The effects $f_\mu(z_i)$ and $f_\sigma(z_i)$ are specified by:
\begin{enumerate}[label=(\alph*)]
\setlength\itemsep{-0.2em}
\item categorical categorical effect 
\begin{align*}
f_\mu(z_i) =
\begin{cases}
-2  &\text{if } z_i = 1 \\
-1.5  &\text{if } z_i = 2 \\
0  &\text{if } z_i = 3 \\
1.5  &\text{if } z_i = 4 \\
2  &\text {if } z_i = 5 \\
\end{cases}
\quad
f_\sigma(z_i) =
\begin{cases}
-0.4  &\text{if } z_i = 1 \\
-0.2  &\text{if } z_i = 2 \\
0  &\text{if } z_i = 3 \\
0.2  &\text{if } z_i = 4 \\
0.4  &\text {if } z_i = 5 \\
\end{cases} ,
\end{align*}
where $Z_i \sim \mathcal{U}\{1,...,5\}$
\item non-linear effect
\begin{align*}
f_\mu(z_i) = 8 \sin\left(\frac{\pi}{2}+z_i\right)-6.5
\quad
f_\sigma(z_i) = \frac{1}{6} \left(2.85 z_i\right)^3 - 2.85 z_i , 
\end{align*}
where $Z_i \sim \mathcal{U}(-1,1)$
\item discrete spatial effect: \\[0.3em]
We consider the neighborhood structure in Nigeria 
with 6 regions $z_i \in \{1, ..., 6\}$.
The continuous spatial effect
\begin{align*}
f(s_1, s_2) = s_1 + s_2
\end{align*}
is averaged over each region and the resulting regional effects are centered around 0 and scaled to the chosen effect size. 
This results in a spatial effect of approximately 
\begin{align*}
f_\mu(z_i) =
\begin{cases}
-0.01 &\text{if } z_i = 1 \\
3  &\text{if } z_i = 2 \\
1.41  &\text{if } z_i = 3 \\
-1.03  &\text{if } z_i = 4 \\
-1.45  &\text {if } z_i = 5 \\
-1.91  &\text {if } z_i = 9 \\
\end{cases}
\quad
f_\sigma(z_i) =
\begin{cases}
-0.001  &\text{if } z_i = 1 \\
0.3  &\text{if } z_i = 2 \\
0.141  &\text{if } z_i = 3 \\
-0.103  &\text{if } z_i = 4 \\
-0.145  &\text {if } z_i = 5 \\
-0.191  &\text {if } z_i = 6 \\
\end{cases} ,
\end{align*}
where $Z_i \sim \mathcal{U}\{1,...,6\}$.
\end{enumerate}

\subsection{ZINB setting}
\label{daub:appdx_zinb_setting}
The effects $f_\mu(z_i)$ and $f_\sigma(z_i)$ are specified by:
\begin{enumerate}[label=(\alph*)]
\setlength\itemsep{-0.2em}
\item categorical categorical effect 
\begin{align*}
f_\mu(z_i) =
\begin{cases}
0.2  &\text{if } z_i = 1 \\
0.1  &\text{if } z_i = 2 \\
0  &\text{if } z_i = 3 \\
-0.1  &\text{if } z_i = 4 \\
-0.2  &\text {if } z_i = 5 \\
\end{cases}
\quad
f_\alpha(z_i) =
\begin{cases}
0.25  &\text{if } z_i = 1 \\
0.15  &\text{if } z_i = 2 \\
0  &\text{if } z_i = 3 \\
-0.15  &\text{if } z_i = 4 \\
-0.25  &\text {if } z_i = 5 \\
\end{cases} 
\quad
f_\pi(z_i) =
\begin{cases}
0.8  &\text{if } z_i = 1 \\
0.6  &\text{if } z_i = 2 \\
0  &\text{if } z_i = 3 \\
-0.6  &\text{if } z_i = 4 \\
-0.8  &\text {if } z_i = 5 \\
\end{cases} ,
\end{align*}
where $Z_i \sim \mathcal{U}\{1,...,5\}$
\item non-linear effect
\begin{align*}
f_\mu(z_i) &= -0.7(\ln(z_i + 1.15)-0.9 z_i) - 0.03)
\quad
f_\alpha(z_i) = 2.1 \text{sin}\left(\frac{\pi}{2}+z_i \right) - 1.767 \\
f_\pi(z_i) &= \frac{1}{3}(2 z_i)^3 - 2 z_i , 
\end{align*}
            
where $Z_i \sim \mathcal{U}(-1,1)$
\item discrete spatial effect \\[0.3em]
We consider the neighborhood structure in Nigeria 
with 6 regions $z_i \in \{1, ..., 6\}$.
The continuous spatial effect
\begin{align*}
f(s_1, s_2) = s_1 + s_2
\end{align*}
is averaged over each region and the resulting regional effects are centered around 0 and scaled to the chosen effect size. 
This yields approximately
\begin{align*}
f_\mu(z_i) =
\begin{cases}
0.001  &\text{if } z_i = 1 \\
-0.3  &\text{if } z_i = 2 \\
-0.141  &\text{if } z_i = 3 \\
0.103  &\text{if } z_i = 4 \\
0.145  &\text {if } z_i = 5 \\
0.191  &\text {if } z_i = 6 \\
\end{cases}
\quad
f_\alpha(z_i) =
\begin{cases}
-0.001  &\text{if } z_i = 1 \\
0.3  &\text{if } z_i = 2 \\
0.141  &\text{if } z_i = 3 \\
-0.103  &\text{if } z_i = 4 \\
-0.145  &\text {if } z_i = 5 \\
-0.191  &\text {if } z_i = 6 \\
\end{cases}
\quad
f_\pi(z_i) =
\begin{cases}
-0.001  &\text{if } z_i = 1 \\
0.3  &\text{if } z_i = 2 \\
0.141  &\text{if } z_i = 3 \\
-0.103  &\text{if } z_i = 4 \\
-0.145  &\text {if } z_i = 5 \\
-0.191  &\text {if } z_i = 6 \\
\end{cases},
\end{align*}
where $Z_i \sim \mathcal{U}\{1,...,6\}$.
\end{enumerate}

\subsection{Further Results for the Gaussian Setting}

\begin{figure}[!htb]
\includegraphics[width=\textwidth]{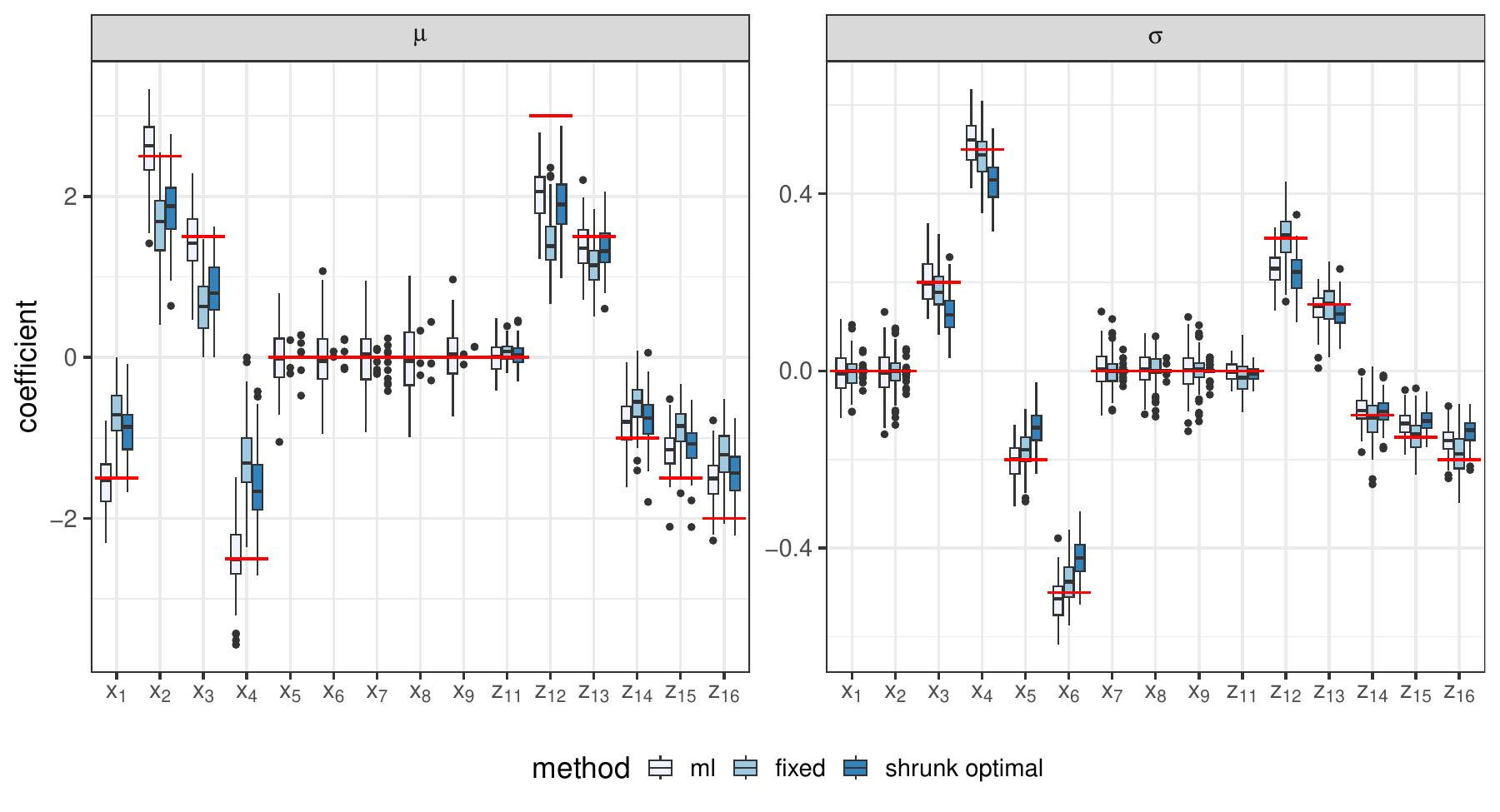}
\caption{Distribution of the coefficient estimates in the Gaussian simulation setting (\ref{daub:simu_gaussian_setting}) with an informative spatial effect. The red horizontal lines represent the true coefficients} 
\end{figure}

\begin{table}[!htb]
\caption{Number of selections of the covariate effects in the Gaussian simulation setting (\ref{daub:simu_gaussian_setting}) with informative spatial effect for 100 simulation runs. The informative effects are marked in bold }
\tabcolsep=0pt
\begin{tabular*}{\textwidth}{@{\extracolsep{\fill}}lcccc@{\extracolsep{\fill}}}
\toprule%
& \multicolumn{2}{@{}c@{}}{$\eta_\mu$} & \multicolumn{2}{@{}c@{}}{$\eta_\sigma$} \\
\cmidrule{2-3}\cmidrule{4-5}%
& fixed & opt & fixed & opt \\ 
\midrule
$x_1$ & \textbf{97} & \textbf{100} & 83 & 12 \\ 
  $x_2$ & \textbf{100} & \textbf{100} & 86 & 14 \\ 
  $x_3$ & \textbf{91} & \textbf{98} & \textbf{100} & \textbf{100} \\ 
  $x_4$ & \textbf{99} & \textbf{100} & \textbf{100} & \textbf{100} \\ 
  $x_5$ & 3 & 7 & \textbf{100} & \textbf{100} \\ 
  $x_6$ & 2 & 5 & \textbf{100} & \textbf{100} \\ 
  $x_7$ & 6 & 10 & 76 & 13 \\ 
  $x_8$ & 3 & 4 & 71 & 4 \\ 
  $x_9$ & 2 & 2 & 88 & 13 \\ 
  $x_{10}$ & 4 & 6 & 84 & 10 \\ 
  $z_\text{inf}$ & \textbf{100} & \textbf{100} & \textbf{100} & \textbf{100} \\ 
  $z_\text{non-inf}$ & -- & -- & -- & -- \\ 
\bottomrule
\end{tabular*}
\end{table}

\begin{figure}[!htb]
\includegraphics[width=\textwidth]{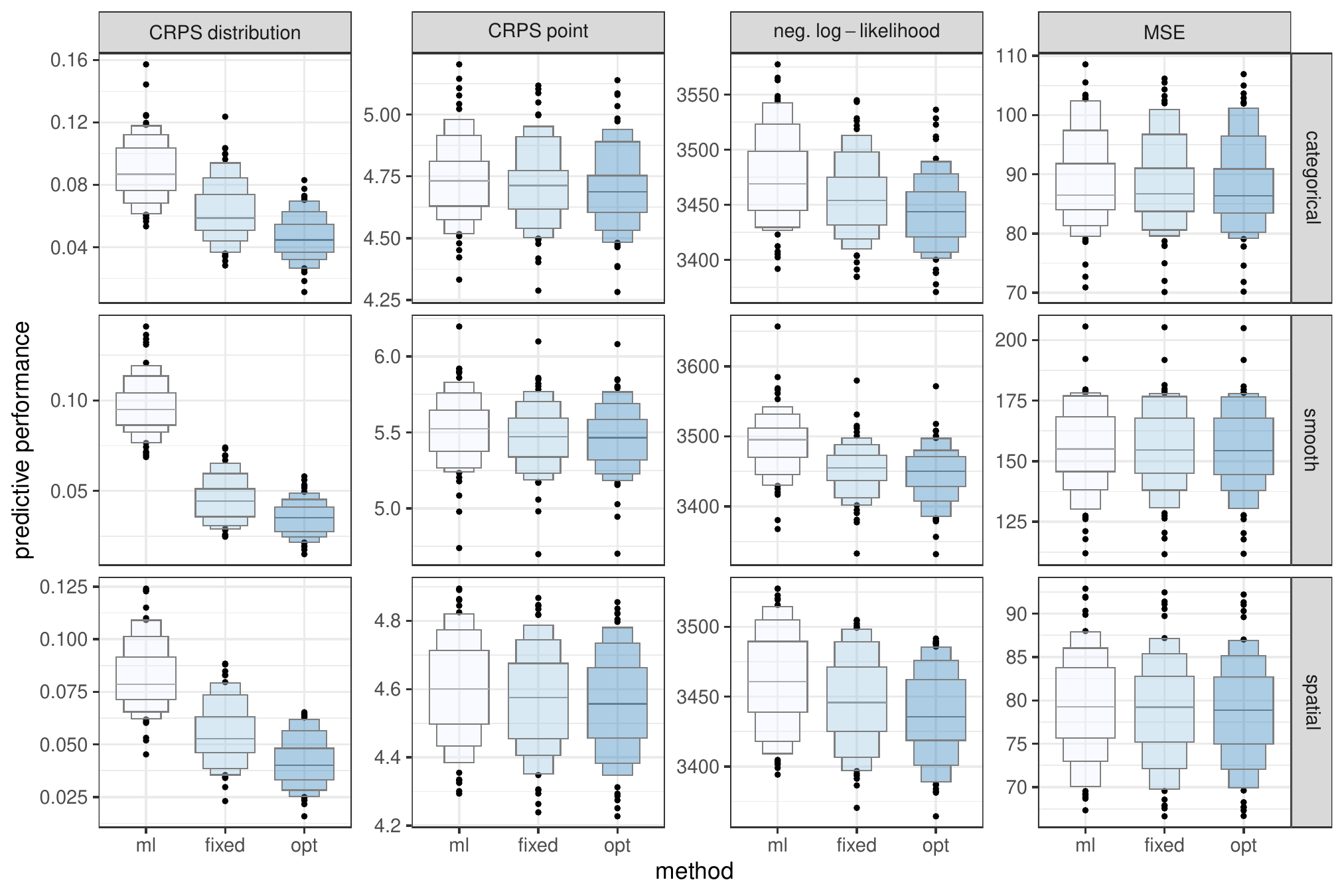}
\caption{Distribution of different predictive performance measures (columns) in the Gaussian simulation setting (\ref{daub:simu_gaussian_setting}) with different additional effects (rows) } 
\end{figure}

\clearpage

\subsection{Further Results for the ZINB Setting}

\begin{figure}[!htb]
\centering
\includegraphics[width=\textwidth]{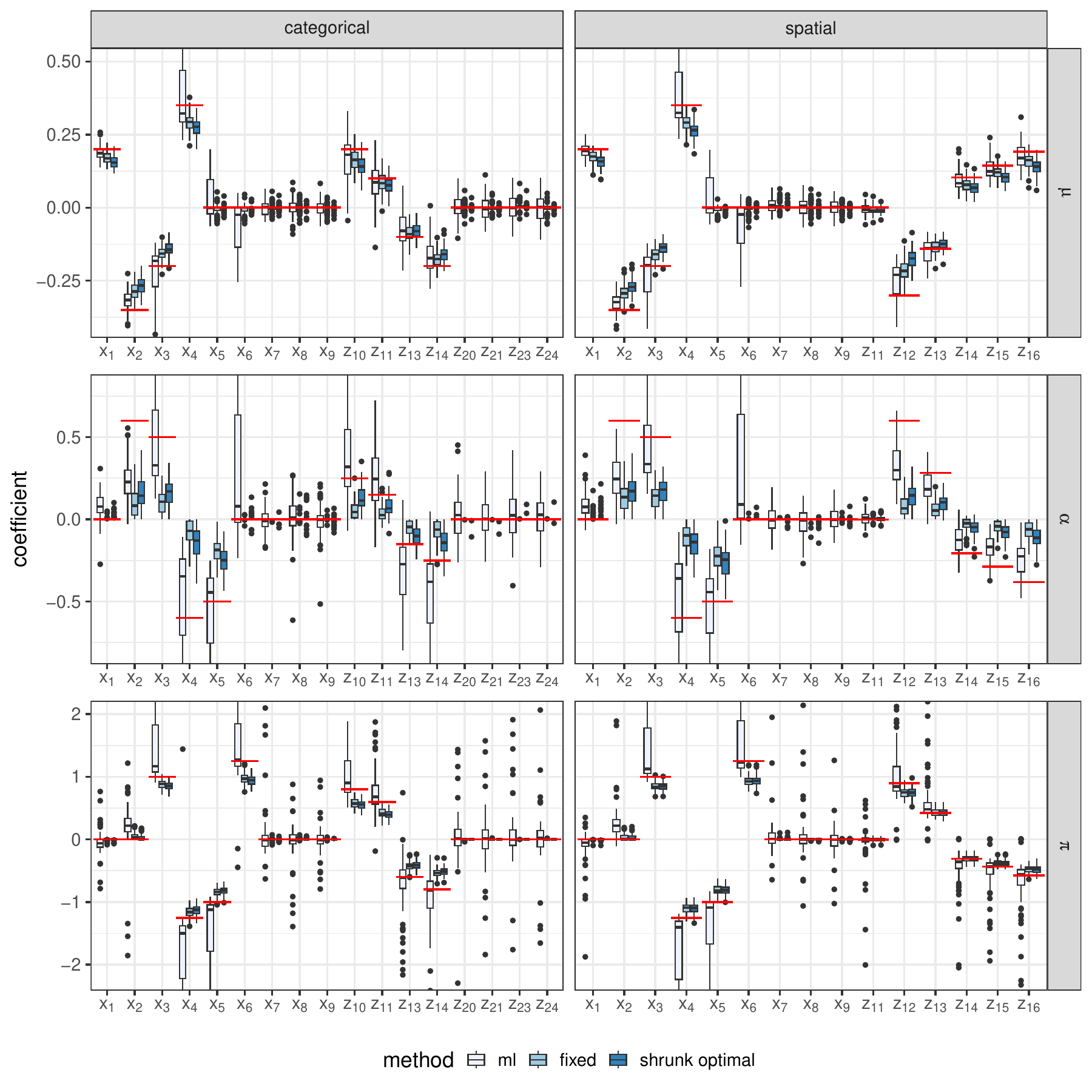}
\caption{Distribution of the coefficient estimates in the ZINB simulation setting (\ref{daub:simu_ZINB_setting}) with categorical effects (left) as well as an informative spatial effect (right). The red horizontal lines represent the true coefficients} 
\end{figure}

%

\begin{figure}[!htb]
\includegraphics[width=\textwidth]{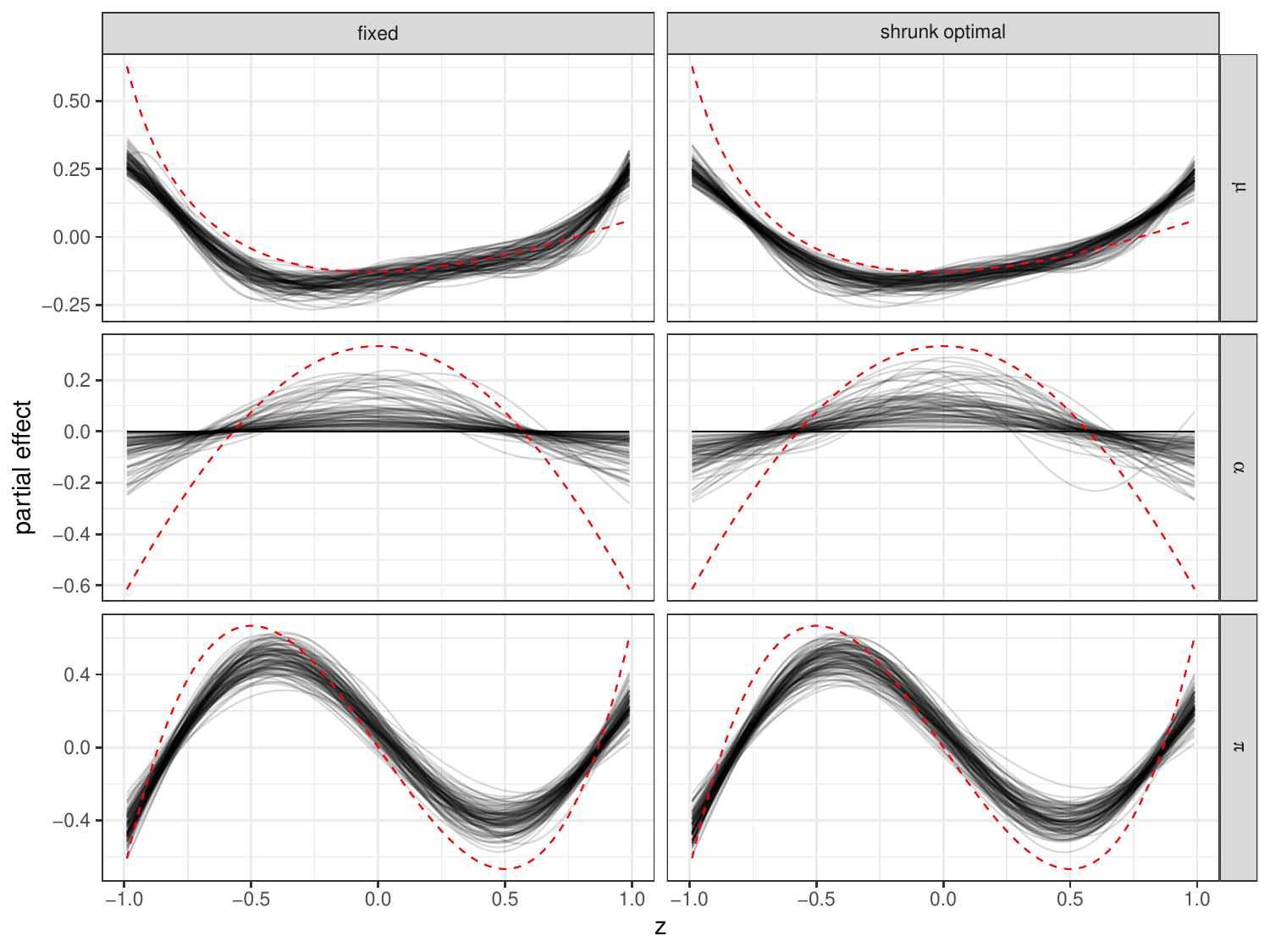}
\caption{Partial effects of the informative non-linear effect in the ZINB simulation setting (\ref{daub:simu_ZINB_setting}). The red dashed lines represent the true partial effect }
\end{figure}

\begin{table}[!t]
\caption{Number of selections of the covariate effects in the ZINB simulation setting (\ref{daub:simu_ZINB_setting}) with additional informative (columns 1-6) and non-informative spatial effect (columns 7-12)  for 100 simulation runs. The informative effects are marked in bold}
\tabcolsep=0pt
\begin{tabular*}{\textwidth}{@{\extracolsep{\fill}}lcccccccccccc@{\extracolsep{\fill}}}
\toprule%
& \multicolumn{6}{@{}c@{}}{categorical} & \multicolumn{6}{@{}c@{}}{non-linear} \\
\cmidrule{2-7}\cmidrule{8-13}%
& \multicolumn{2}{@{}c@{}}{$\eta_\mu$} & \multicolumn{2}{@{}c@{}}{$\eta_\alpha$} & \multicolumn{2}{@{}c@{}}{$\eta_\pi$} & \multicolumn{2}{@{}c@{}}{$\eta_\mu$} & \multicolumn{2}{@{}c@{}}{$\eta_\alpha$} & \multicolumn{2}{@{}c@{}}{$\eta_\pi$} \\
\cmidrule{2-3}\cmidrule{4-5}\cmidrule{6-7}\cmidrule{8-9}\cmidrule{10-11}\cmidrule{12-13}%
& fixed & opt & fixed & opt & fixed & opt & fixed & opt & fixed & opt & fixed & opt \\ 
\midrule
$x_1$ & \textbf{100} & \textbf{100} & 12 & 13 & 5 & 5 & \textbf{100} & \textbf{100} & 5 & 10 & 4 & 3 \\ 
  $x_2$ & \textbf{100} & \textbf{100} & \textbf{87} & \textbf{95} & 46 & 45 & \textbf{100} & \textbf{100} & \textbf{87} & \textbf{93} & 36 & 33 \\ 
  $x_3$ & \textbf{100} & \textbf{100} & \textbf{92} & \textbf{92} & \textbf{100} & \textbf{100} & \textbf{100} & \textbf{100} & \textbf{93} & \textbf{96} & \textbf{100} & \textbf{100} \\ 
  $x_4$ & \textbf{100} & \textbf{100} & \textbf{88} & \textbf{95} & \textbf{100} & \textbf{100} & \textbf{100} & \textbf{100} & \textbf{84} & \textbf{96} & \textbf{100} & \textbf{100} \\ 
  $x_5$ & 45 & 19 & \textbf{100} & \textbf{100} & \textbf{100} & \textbf{100} & 52 & 11 & \textbf{98} & \textbf{99} & \textbf{100} & \textbf{100} \\ 
  $x_6$ & 45 & 12 & 2 & 3 & \textbf{100} & \textbf{100} & 33 & 9 & 2 & 5 & \textbf{100} & \textbf{100} \\ 
  $x_7$ & 51 & 15 & 1 & 7 & 4 & 3 & 25 & 3 & 2 & 2 & 1 & 1 \\ 
  $x_8$ & 56 & 16 & 3 & 8 & 2 & 3 & 25 & 1 & 0 & 3 & 1 & 0 \\ 
  $x_9$ & 54 & 16 & 4 & 6 & 4 & 5 & 29 & 4 & 0 & 3 & 1 & 1 \\ 
  $x_{10}$ & 45 & 10 & 3 & 8 & 4 & 4 & 35 & 6 & 0 & 6 & 3 & 3 \\ 
  $z_1$ & \textbf{100} & \textbf{100} & \textbf{82} & \textbf{96} & \textbf{100} & \textbf{100} & -- & -- & -- & -- & -- & -- \\ 
  $z_2$ & -- & -- & -- & -- & -- & -- & 64 & 11 & 0 & 6 & 1 & 2 \\ 
\bottomrule
\end{tabular*}
\end{table}

\begin{figure}[!htb]
\includegraphics[width=\textwidth]{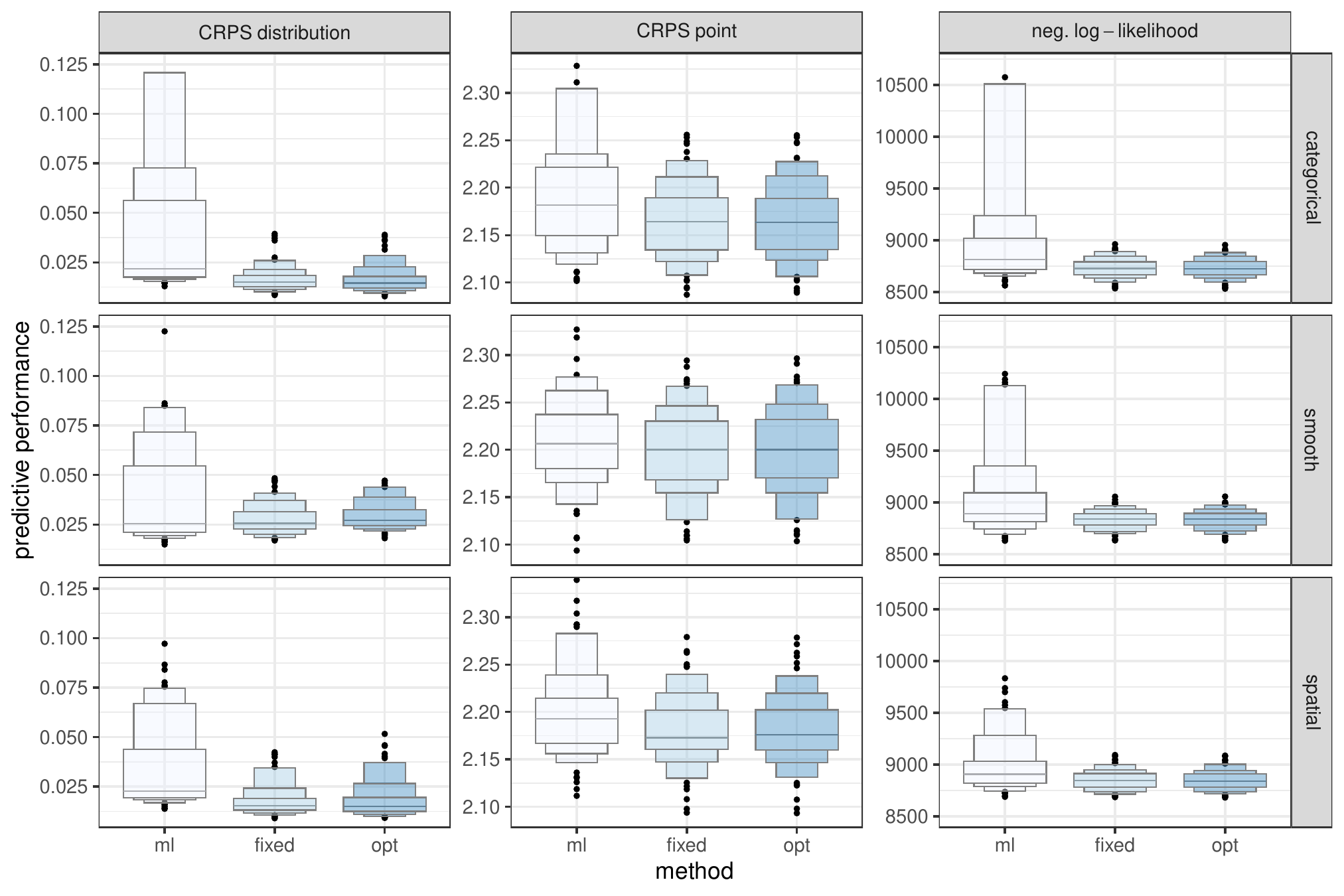}
\caption{
Distribution of different predictive performance measures (columns) in the ZINB simulation setting (\ref{daub:simu_ZINB_setting}) with different additional effects (rows). For better visibility, some outliers of the maximum likelihood method are excluded }
\end{figure}

%

\clearpage

\section{Application}
\label{daub:appdx_application}

\subsection{Information on the Data}

\begin{table}[!htb]
\caption{Overview of the variables and their summary statistics for the antenatal care data from Nigeria 
\label{daub:table_descriptives_antenal_care}}
\begin{tabular*}{\columnwidth}{@{\extracolsep\fill}llcccc@{\extracolsep\fill}}
\toprule
variable & short description & mean/frequency & standard dev. & min & max \\ 
\midrule
\textbf{antenatal} & number of antenatal care visits & 5.25 & 6.11 & 0 & 30 \\ 
\textbf{age} & age in years & 28.46 & 7.08 & 13 & 50 \\ 
\textbf{married\_yrs} & marriage duration in years & 11.63 & 7.51 & 0 & 39 \\ 
\textbf{education} & 0 = no education & 47.1\% \\ 
& 1 = primary school education & 20.4\% \\ 
& 2 = higher education & 32.5\% \\ 
\textbf{wealth} & 0 = poorest  & 22.4\% \\ 
& 1 = poor  & 23.1\% \\ 
& 2 = middle  & 19.8\% \\ 
& 3 = richer  & 18.4\% \\ 
& 4 = richest  & 16.2\% \\ 
\textbf{work} & 1 = employed & 69.4\% \\ 
& 0 = unemployed  & 30.6 \% \\ 
\textbf{television} & 1 = access to television & 45.8\% \\ 
& 0 = no access  & 54.2\% \\ 
\textbf{radio} & 1 = access to radio & 59.1\% \\ 
& 0 = no access  & 40.9\% \\ 
\textbf{newspaper} & 1 = access to newspaper & 14.3\% \\ 
& 0 = no access  & 85.7\% \\ 
\textbf{urban} & 1 = urban area of residence & 33.1\% \\
& 0 = rural area of residence  & 66.9\% \\ 
\textbf{zone} & 221 = North Central  & 15.5\% \\ 
& 222 = North East  & 20.1\% \\ 
& 223 = North West  & 32.0\% \\ 
& 224 = South East  & 8.1\% \\ 
& 225 = South South  & 11.1\% \\ 
& 226 = South West  & 13.2\% \\ 
\bottomrule
\end{tabular*}
\end{table}

\clearpage

\subsection{Further Results}

\begin{table}[!htb]
\caption{Selected coefficients for the antenatal care data from Nigeria using different step length approaches. The shrunk optimal approach is referred to by ``opt'' \label{daub:table_antenatal_coef_all}}
\tabcolsep=0pt
\begin{tabular*}{\textwidth}{@{\extracolsep{\fill}}lcccccc@{\extracolsep{\fill}}}
\toprule%
& \multicolumn{2}{@{}c@{}}{$\eta_\mu$} & \multicolumn{2}{@{}c@{}}{$\eta_\alpha$} & \multicolumn{2}{@{}c@{}}{$\eta_\pi$} \\
\cmidrule{2-3}\cmidrule{4-5}\cmidrule{6-7}%
& fixed & opt & fixed & opt & fixed & opt \\ 
\midrule
urban & 0.016 & 0.000 & -0.192 & -0.204 & -0.619 & -0.641 \\ 
  newsp & 0.020 & 0.025 & -0.266 & -0.258 & -0.143 & -0.140 \\ 
  radio & 0.048 & 0.034 & -0.051 & -0.073 & -0.175 & -0.170 \\ 
  telv & 0.047 & 0.058 & -0.033 & -0.041 & -0.472 & -0.482 \\ 
  work & 0.000 & 0.000 & 0.000 & 0.000 & -0.218 & -0.208 \\ 
  education0 & -0.033 & -0.027 & 0.000 & 0.000 & 0.603 & 0.594 \\ 
  education2 & 0.019 & 0.020 & 0.000 & 0.000 & -0.457 & -0.450 \\ 
  wealth0 & -0.033 & -0.028 & 0.013 & 0.033 & 0.519 & 0.497 \\ 
  wealth1 & -0.041 & -0.037 & -0.007 & -0.002 & 0.408 & 0.382 \\ 
  wealth3 & 0.028 & 0.025 & -0.034 & -0.058 & -0.298 & -0.289 \\ 
  wealth4 & 0.033 & 0.031 & -0.053 & -0.089 & -0.517 & -0.479 \\ 
  urban:radio & -0.015 & 0.000 & 0.000 & 0.000 & 0.000 & 0.000 \\ 
  urban:telv & -0.007 & 0.000 & 0.000 & 0.000 & 0.000 & 0.000 \\ 
  urban:work & 0.006 & 0.006 & 0.000 & 0.000 & 0.000 & 0.000 \\ 
  urban:wealth0 & -0.096 & -0.048 & 0.000 & 0.000 & 0.039 & 0.038 \\ 
  urban:wealth1 & -0.043 & -0.025 & 0.000 & 0.000 & 0.000 & 0.000 \\ 
  urban:wealth3 & -0.009 & 0.000 & 0.000 & 0.000 & 0.000 & 0.000 \\ 
  urban:wealth4 & 0.022 & 0.023 & 0.000 & 0.000 & -0.117 & -0.051 \\ 
  urban:education0 & -0.005 & 0.000 & 0.000 & 0.000 & 0.000 & 0.000 \\ 
  urban:education2 & 0.042 & 0.046 & 0.000 & 0.000 & 0.000 & 0.000 \\ 
  newsp:work & 0.021 & 0.010 & 0.000 & 0.000 & 0.000 & 0.000 \\ 
  newsp:wealth0 & 0.092 & 0.000 & 0.000 & 0.000 & 0.000 & 0.000 \\ 
  newsp:wealth1 & 0.013 & 0.000 & 0.000 & 0.000 & 0.000 & 0.000 \\ 
  newsp:education0 & -0.105 & 0.000 & 0.000 & 0.000 & 0.000 & 0.000 \\ 
  radio:wealth0 & 0.000 & 0.000 & 0.000 & 0.000 & 0.023 & 0.015 \\ 
  radio:wealth3 & 0.004 & 0.000 & 0.000 & 0.000 & 0.000 & 0.000 \\ 
  radio:wealth4 & 0.000 & 0.000 & -0.023 & -0.034 & -0.067 & -0.093 \\ 
  radio:education2 & 0.000 & 0.016 & 0.000 & 0.000 & 0.000 & 0.000 \\ 
  telv:wealth1 & -0.022 & -0.013 & 0.000 & 0.000 & 0.000 & 0.000 \\ 
  telv:wealth4 & 0.031 & 0.022 & 0.000 & 0.000 & 0.000 & 0.000 \\ 
  telv:education0 & -0.005 & 0.000 & 0.000 & 0.000 & 0.000 & 0.000 \\ 
  telv:education2 & 0.040 & 0.025 & 0.000 & 0.000 & 0.000 & 0.000 \\ 
  work:wealth0 & -0.028 & 0.000 & 0.000 & 0.000 & 0.000 & 0.000 \\ 
  work:wealth1 & -0.018 & -0.004 & 0.000 & 0.000 & 0.000 & 0.000 \\ 
  work:wealth3 & 0.021 & 0.013 & 0.000 & 0.000 & 0.000 & 0.000 \\ 
  work:wealth4 & 0.000 & 0.008 & 0.000 & 0.000 & 0.000 & 0.000 \\ 
  wealth0:education0 & 0.000 & 0.000 & 0.000 & 0.000 & 0.563 & 0.552 \\ 
  wealth0:education2 & -0.038 & 0.000 & 0.000 & 0.000 & 0.000 & 0.000 \\ 
  wealth1:education0 & 0.000 & 0.000 & 0.000 & 0.000 & 0.115 & 0.118 \\ 
  wealth1:education2 & 0.000 & 0.000 & 0.000 & 0.031 & 0.000 & 0.000 \\ 
  wealth4:education0 & 0.000 & 0.000 & 0.000 & -0.089 & 0.000 & 0.000 \\ 
  zone221 & -0.052 & -0.047 & 0.045 & 0.084 & -0.131 & -0.133 \\ 
  zone222 & -0.356 & -0.372 & -0.583 & -0.679 & -0.286 & -0.284 \\ 
  zone223 & -0.381 & -0.386 & -0.580 & -0.686 & 0.662 & 0.661 \\ 
  zone224 & 0.157 & 0.147 & 0.291 & 0.330 & -0.581 & -0.543 \\ 
  zone225 & 0.085 & 0.116 & 0.499 & 0.567 & 0.772 & 0.722 \\ 
  zone226 & 0.548 & 0.542 & 0.327 & 0.384 & -0.436 & -0.423 \\  
\bottomrule
\end{tabular*}
\end{table}

\begin{figure}[!htb]
\centering
\includegraphics[width=\textwidth]{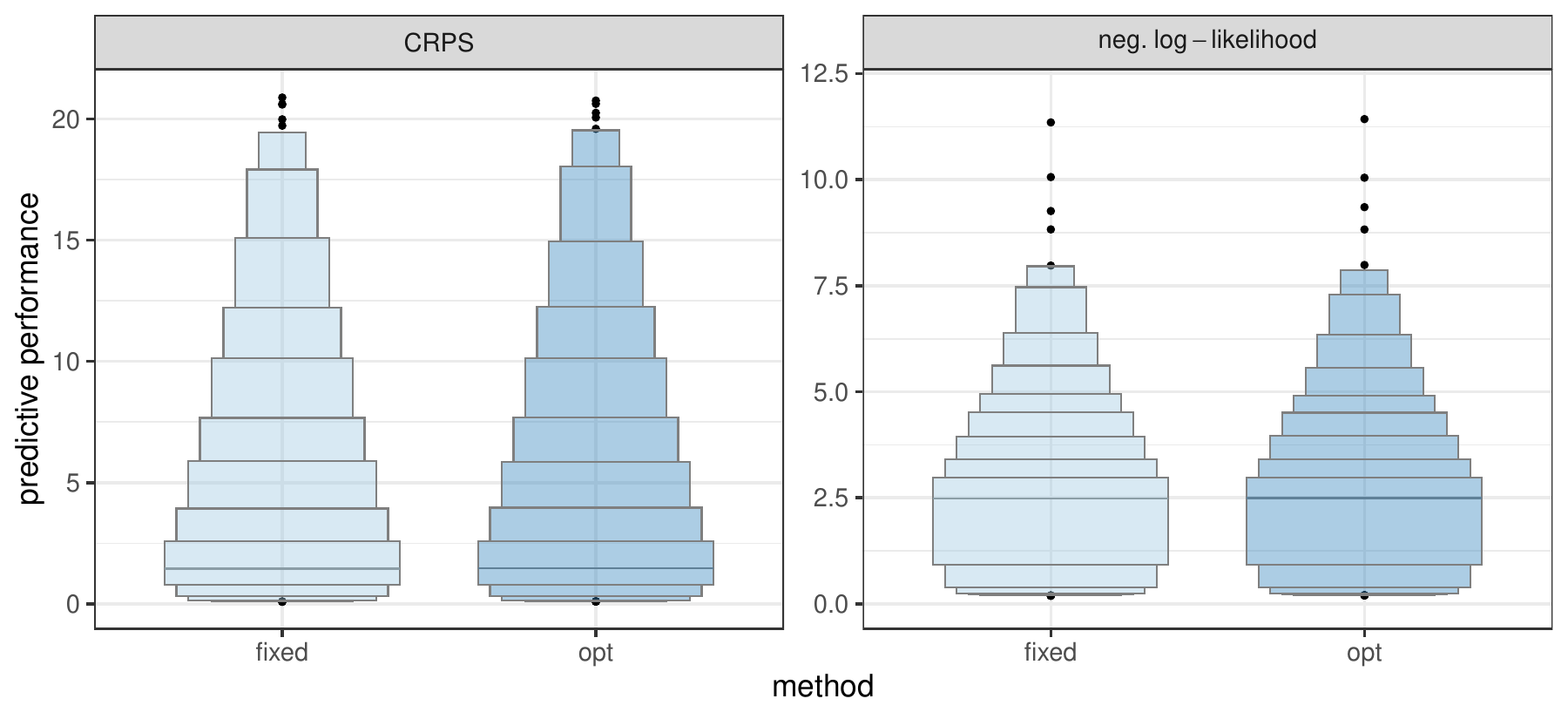}
\caption{Distribution of different predictive performance measures applied on each observation for the validation set of the antenatal care data from Nigeria (columns) using different step length approaches. For better visibility, few outliers are excluded}
\end{figure}

\clearpage

\subsection{Comparison of Boosting Approaches with Direct Maximum Likelihood Estimation without Interactions}

\begin{table}[!htb]
\caption{Coefficients for the antenatal care data from Nigeria without interactions  using different step length approaches. ``ml'' refers to the direct maximum likelihood estimation via the \textit{gamlss} package and the shrunk optimal approach is abbreviated by ``opt'' }
\tabcolsep=0pt
\begin{tabular*}{\textwidth}{@{\extracolsep{\fill}}lccccccccc@{\extracolsep{\fill}}}
\toprule%
& \multicolumn{3}{@{}c@{}}{$\eta_\mu$} & \multicolumn{3}{@{}c@{}}{$\eta_\alpha$} & \multicolumn{3}{@{}c@{}}{$\eta_\pi$} \\
\cmidrule{2-4}\cmidrule{5-7}\cmidrule{8-10}%
& ml & fixed & opt & ml & fixed & opt & ml & fixed & opt \\ 
\midrule
urban & -0.005 & 0.029 & 0.032 & -0.093 & -0.188 & -0.202 & -0.667 & -0.611 & -0.630 \\ 
  newsp & 0.020 & 0.039 & 0.034 & -0.280 & -0.268 & -0.256 & -0.461 & -0.115 & -0.119 \\ 
  radio & 0.052 & 0.033 & 0.035 & -0.071 & -0.045 & -0.058 & -0.249 & -0.176 & -0.179 \\ 
  telv & 0.039 & 0.076 & 0.066 & 0.014 & -0.025 & -0.045 & -0.263 & -0.444 & -0.441 \\ 
  work & 0.017 & 0.000 & 0.000 & 0.081 & 0.000 & 0.000 & -0.325 & -0.225 & -0.230 \\ 
  education0 & -0.054 & -0.047 & -0.044 & -0.249 & 0.000 & 0.000 & 0.878 & 0.741 & 0.741 \\ 
  education2 & 0.068 & 0.058 & 0.057 & 0.072 & 0.000 & 0.000 & -0.325 & -0.503 & -0.503 \\ 
  wealth0 & -0.087 & -0.048 & -0.037 & 0.176 & 0.018 & 0.028 & 1.151 & 0.972 & 0.971 \\ 
  wealth1 & -0.080 & -0.060 & -0.052 & -0.041 & -0.005 & -0.002 & 0.620 & 0.445 & 0.443 \\ 
  wealth3 & 0.064 & 0.033 & 0.023 & -0.202 & -0.040 & -0.050 & -0.305 & -0.368 & -0.368 \\ 
  wealth4 & 0.117 & 0.079 & 0.070 & -0.293 & -0.071 & -0.089 & -1.453 & -0.668 & -0.659 \\ 
  zone221 & -0.055 & -0.052 & -0.047 & 0.077 & 0.048 & 0.074 & -0.105 & -0.148 & -0.149 \\ 
  zone222 & -0.317 & -0.352 & -0.371 & -0.563 & -0.598 & -0.655 & -0.435 & -0.301 & -0.312 \\ 
  zone223 & -0.346 & -0.375 & -0.384 & -0.557 & -0.595 & -0.656 & 0.403 & 0.658 & 0.650 \\ 
  zone224 & 0.145 & 0.147 & 0.141 & 0.233 & 0.307 & 0.325 & -0.333 & -0.622 & -0.618 \\ 
  zone225 & 0.047 & 0.079 & 0.115 & 0.490 & 0.512 & 0.555 & 0.656 & 0.837 & 0.852 \\ 
  zone226 & 0.527 & 0.552 & 0.547 & 0.319 & 0.326 & 0.357 & -0.185 & -0.425 & -0.423 \\  
\bottomrule
\end{tabular*}
\end{table}

\begin{figure}[!htb]
\centering
\includegraphics[width=0.8\textwidth]{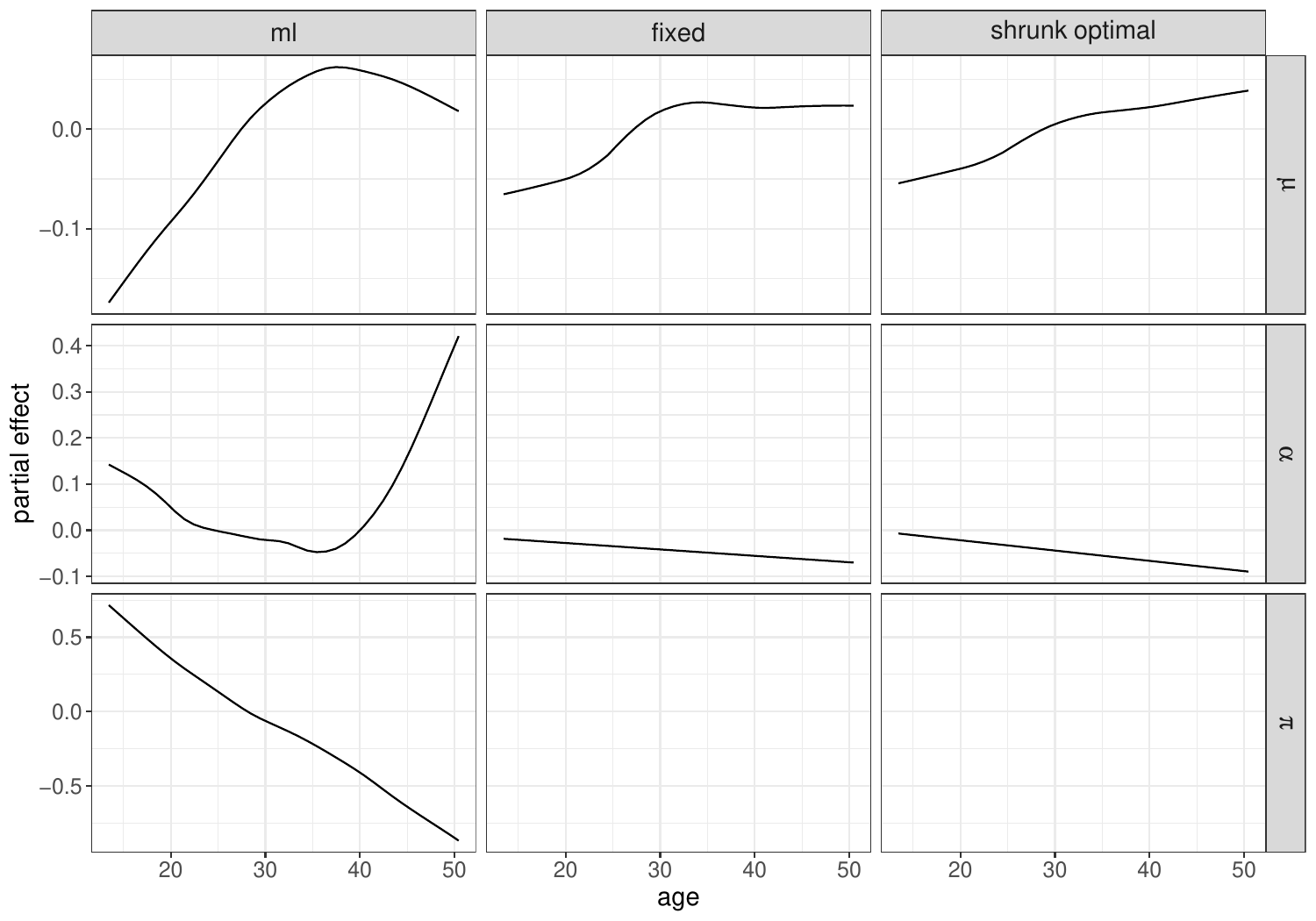}
\caption{Partial effects of the mother's age for the antenatal care data from Nigeria modeled without interactions using different step length approaches (columns). ``ml'' refers to the direct maximum likelihood estimation via the \textit{gamlss} package}
\end{figure}

\begin{figure}[!htb]
\centering
\includegraphics[width=0.8\textwidth]{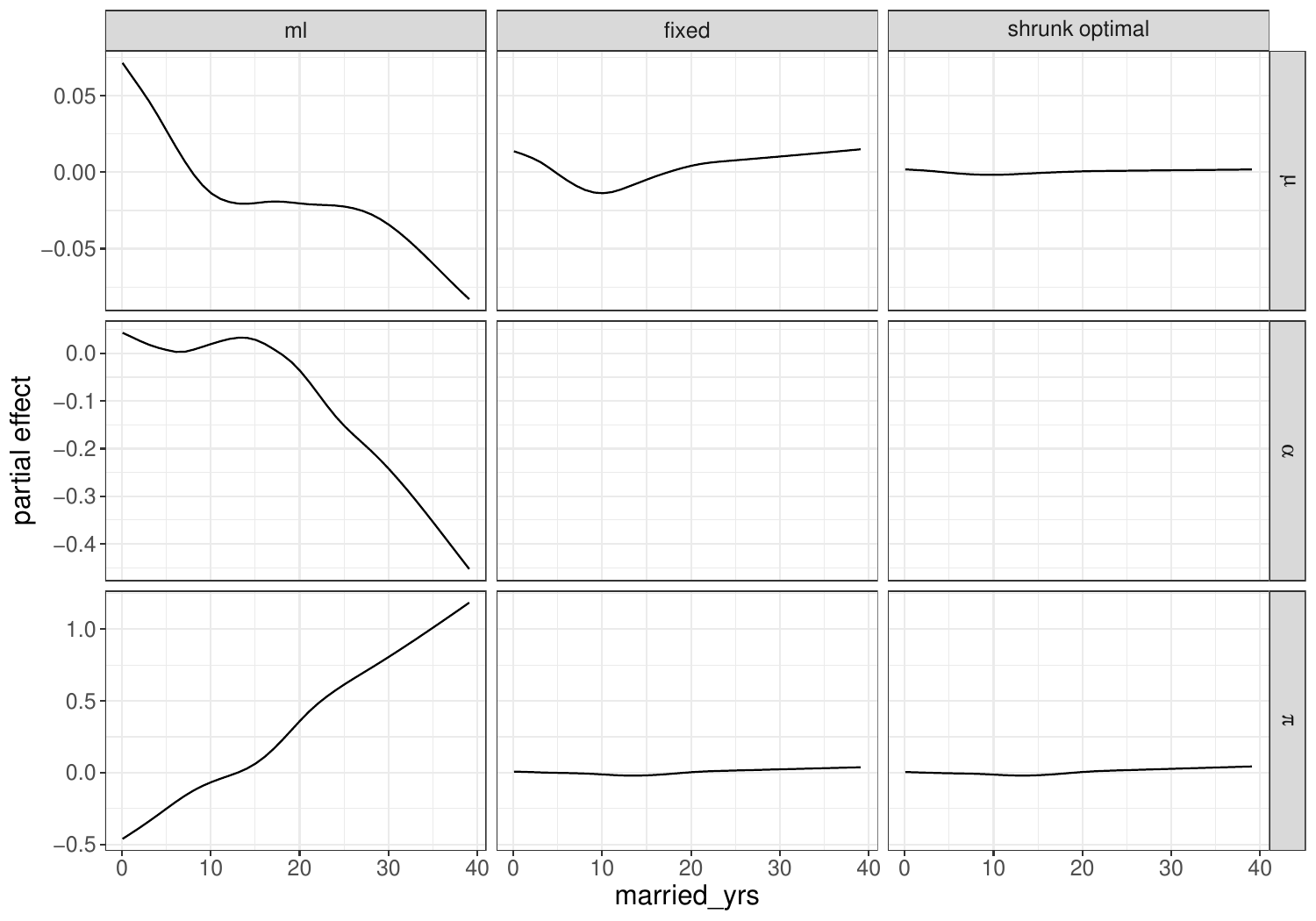}
\caption{Partial effects of the marriage duration for the antenatal care data from Nigeria modeled without interactions using different step length approaches (columns). ``ml'' refers to the direct maximum likelihood estimation via the \textit{gamlss} package}
\end{figure}

\begin{figure}[!htb]
\centering
\includegraphics[width=0.9\textwidth]{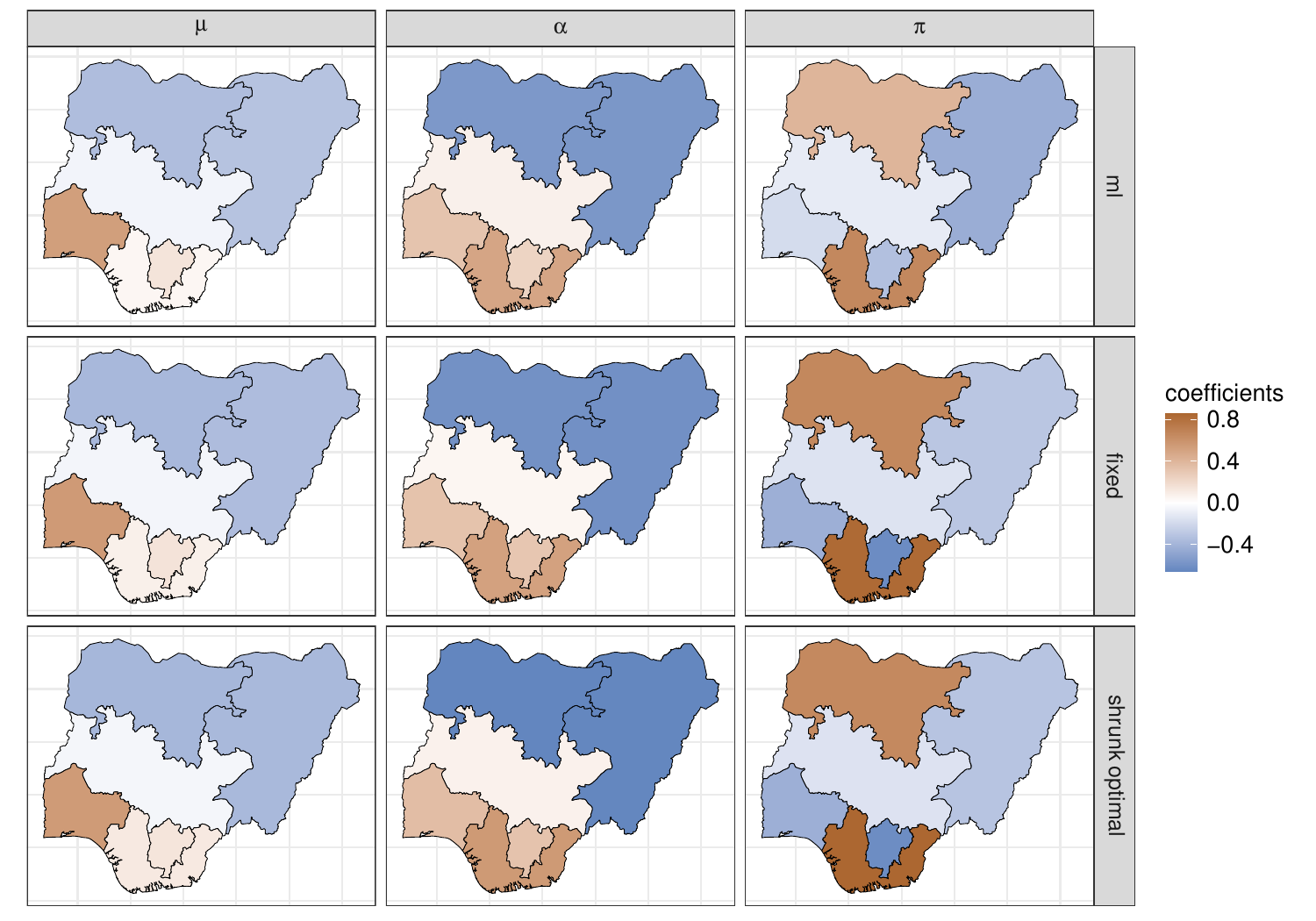}
\caption{Spatial effects for the antenatal care data from Nigeria modeled without interactions using different step length approaches (rows). ``ml'' refers to the direct maximum likelihood estimation via the \textit{gamlss} package}
\end{figure}

\end{document}